\documentclass[a4paper,11pt]{article}
\usepackage{jheppub}
\usepackage[utf8x]{inputenc}
\usepackage{graphicx}
\usepackage{float}
\usepackage[T1]{fontenc}
\usepackage{epsfig}
\usepackage{color}
\usepackage{amsmath,mathtools}
\usepackage{subfloat}
\usepackage{amsfonts}
\usepackage{braket}
\usepackage{cleveref}
\usepackage{epstopdf}
\usepackage{caption}
\usepackage{subcaption}
\usepackage{enumitem}
\usepackage{comment}
\usepackage[titletoc,toc,title]{appendix}
\usepackage{hyperref}
\hypersetup{
	colorlinks=true,
	linkcolor=blue,
	filecolor=red,      
	urlcolor=blue,
	citecolor=red
}

\DeclareMathOperator{\sech}{sech}
\DeclareMathOperator{\csch}{csch}
\DeclareMathOperator{\arctanh}{arctanh}

\def\be{\begin{equation}}
	\def\ee{\end{equation}}
\def\bea{\begin{eqnarray}}
	\def\eea{\end{eqnarray}}
\def\ba{\begin{align}}
	\def\ea{\end{align}}
\def\d{\text{d}}

\def\d{\text{d}}

\def\d{\mathrm{d}}
\def\cM{\mathcal{M}}
\def\cN{\mathcal{N}}

\def\cL{\mathcal{L}}
\def\cQ{\mathcal{Q}}

\title{Reflected entropy and islands in a braneworld cosmology}

\author[]{Debarshi Basu,$^{a}$}
\author[]{Ashish Chandra,$^{a}$}
\author[]{Himanshu Chourasiya,$^{b}$}

\affiliation[]{
	$^a$Shing-Tung Yau Center and School of Physics, Southeast University,\\
	Nanjing 210096, China\\
	$^b$Department of Physics, Indian Institute of Technology,\\
	Kanpur 208016, India}
\emailAdd{debarshi.128@gmail.com}
\emailAdd{achandrahep@gmail.com}
\emailAdd{chim@iitk.ac.in}
\date{}

\abstract{
This work investigates the nature of mixed state entanglement and correlation in a braneworld cosmological model, where the bulk geometry is described by an eternal BTZ black hole truncated by an end-of-the-world brane representing a Friedmann-Robertson-Walker (FRW) cosmology. We explore the holographic reflected entropy for both adjacent and disjoint subsystems using the island prescription and the defect extremal surface prescription. In the large central charge limit, we demonstrate that both prescriptions yield an exact agreement. Additionally, we analyze the time evolution of reflected entropy and holographic mutual information, along with an analysis of the geometric Markov gap. Our study provides new insights into the role of quantum extremal surfaces in probing black hole interiors and cosmological spacetimes, with implications for understanding mixed state entanglement and quantum information dynamics in holographic cosmological models.}

\begin{document}

	\maketitle
	
	\section{Introduction}\label{sec:intro}
	The black hole information loss paradox has been a long-standing puzzle in theoretical physics. In its semiclassical treatment, Hawking radiation leads to a monotonically increasing entropy of the radiation, suggesting an eventual loss of information and a breakdown of unitary evolution \cite{Hawking:1976ra,Preskill:1992tc}. However, recent breakthroughs involving quantum extremal surfaces (QES) \cite{Engelhardt:2014gca} and the island formula have provided a resolution to this paradox by demonstrating that the fine-grained entropy of Hawking radiation follows a unitary Page curve \cite{Almheiri:2020cfm,Almheiri:2019hni,Almheiri:2019yqk,Almheiri:2019qdq,Penington:2019npb,Penington:2019kki}. The key insight is that quantum extremal islands, disconnected regions in the black hole interior, contribute to the entropy calculation and lead to an eventual entropy decrease, consistent with unitary quantum mechanics \cite{Almheiri:2019psy,Chen:2019uhq,Hartman:2020swn,Alishahiha:2020qza,Krishnan:2020oun,Geng:2020qvw,Dong:2020uxp,Ling:2020laa,Matsuo:2020ypv,Akal:2020twv,Goto:2020wnk,Deng:2020ent,Geng:2020fxl,Karananas:2020fwx,Anderson:2020vwi,Hashimoto:2020cas,Anegawa:2020ezn,Gautason:2020tmk,Chen:2020hmv,Chen:2020jvn,Chen:2020uac,Geng:2021hlu,Geng:2021iyq,Geng:2021mic,Kawabata:2021hac,Verheijden:2021yrb,Wang:2021woy,Arefeva:2021kfx,Yu:2021cgi,Li:2021mjp,Ghosh:2021axl,Kim:2021gzd,Bhattacharya:2021jrn,Wang:2021mqq,Lu:2021gmv,He:2021mst,Hollowood:2021wkw,Hollowood:2021lsw,Tian:2022pso}. 
	
	A related issue arises while considering the Bekenstein bound, which states that the entropy of a system should be bounded by its area \cite{Bekenstein:1980jp}. In black holes, the inclusion of islands restores this bound, preventing entropy from growing indefinitely. However, a similar problem appears in cosmological spacetimes, where the entanglement entropy of quantum fields in an expanding universe can grow without bound \cite{Bousso:1999xy,Bousso:2002ju}. This motivates the question: Can quantum extremal islands exist in cosmology, providing an analogous regulation of entropy growth? Substantial progress has been made in identifying islands in de Sitter (dS) space and Freedman-Robertson-Walker (FRW) cosmologies \cite{Chen:2020tes,Balasubramanian:2020xqf,Hartman:2020khs,Sybesma:2020fxg,Choudhury:2020hil,Geng:2021wcq,Fallows:2021sge,Aalsma:2021bit,Goswami:2021ksw,Bousso:2022gth,Espindola:2022fqb,Goswami:2022ylc,Aalsma:2022swk,Yadav:2022jib,Ben-Dayan:2022nmb}.

	In this article, we consider a closely related model describing the microstate geometries associated with the (Euclidean) time evolved state of a boundary CFT (BCFT) defined on an interval of Euclidean time \cite{Cooper:2018cmb,Almheiri:2018ijj}
	\begin{align}
		\ket{B_{\tau_0}}=e^{-\tau_0H}\ket{B}=\vcenter{\hbox{\includegraphics[width=3cm]{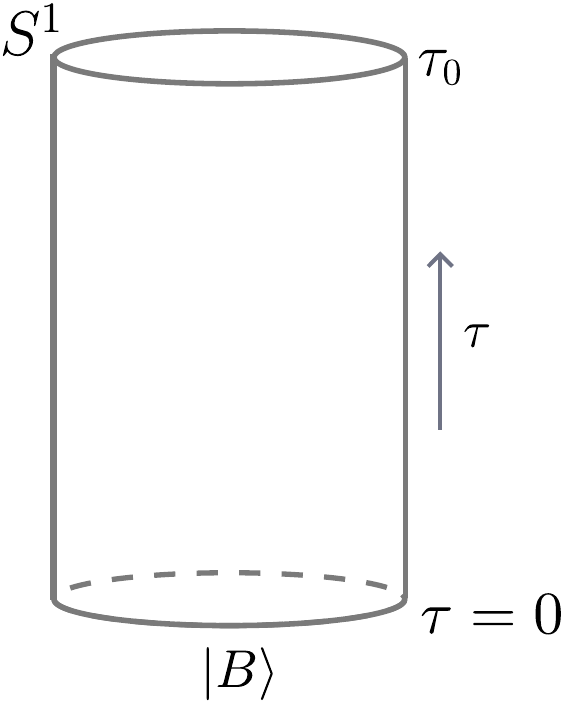}}}
	\end{align}
	This state may be understood as a cousin of the TFD state projected onto some pure state of the left CFT. The bulk dual geometry is described by an eternal BTZ black hole geometry truncated by an end-of-the-world (EOW) brane describing a Freedman-Robertson-Walker cosmology. Recently, a lower dimensional effective description of this state, involving a braneworld cosmology coupled to a thermal bath, was obtained in \cite{Wang:2021afl} through the incorporation of defect matter on the brane and a partial Randall-Sundrum reduction \cite{Randall:1999vf,Randall:1999ee,Karch:2000ct,Deng:2020ent,Chu:2021gdb,Li:2021dmf}. The bulk geometry with defect matter on the EOW brane may then be treated as the so called \textit{doubly holographic} counterpart of the lower dimensional effective theory \cite{Chen:2020uac,Chen:2020hmv,Deng:2020ent,Chu:2021gdb}. Remarkably, in the presence of cosmology of the brane,  \cite{Wang:2021xih} demonstrated the equivalence of the entanglement entropy computed from the defect-extremal-surface (DES) prescription \cite{Deng:2020ent} in the bulk geometry and the island prescription \cite{Almheiri:2020cfm,Almheiri:2019hni,Almheiri:2019qdq,Almheiri:2019psy,Penington:2019npb,Penington:2019kki} in the lower dimensional effective perspective. 
	
	Interestingly, as pointed out in \cite{Cooper:2018cmb}, in this braneworld model, quantum extremal surfaces can probe regions inside black hole horizons, offering a new perspective on black hole interiors. It is then natural to ask whether there exist other finer probes of behind-the-horizon physics. In this context, the mixed state correlation measure termed as the reflected entropy \cite{Dutta:2019gen} offers a natural candidate of such a probe: the quantum extremal cross-section (QECS) or the island cross-section \cite{Chandrasekaran:2020qtn,Li:2020ceg}. From the doubly holographic perspective, the QECS may be identified with the endpoint of the bulk entanglement wedge cross-section (EWCS) \cite{Takayanagi:2017knl,Nguyen:2017yqw} on the EOW brane. Furthermore, the reflected entropy or the EWCS offer a natural framework to analyze the mixed state entanglement and correlation in the braneworld cosmology. Remarkably, a doubly holographic counterpart of the island prescription for the reflected entropy \cite{Chandrasekaran:2020qtn,Li:2020ceg} was put forward in \cite{Li:2021dmf}, in the framework AdS/BCFT with defect conformal matter on the EOW brane. 
	
	In this article, we investigate the mixed state entanglement structure in the braneworld cosmology through the reflected entropy. We obtain reflected entropy and the entanglement wedge cross-section utilizing the bulk defect extremal surface (DES) prescription \cite{Li:2021dmf} for disjoint and adjacent subsystems in the asymptotic boundary and compare our results with the island prescription in the lower dimensional effective theory obtained from partial dimensional reduction. Depending on the relative sizes of the subsystems under consideration, there exists various phases for the reflected entropy and mutual information, indicating a rich phase structure of mixed state entanglement in this defect braneworld theory. Interestingly, we find evidence of extremal surfaces that appear to probe behind-the-horizon physics without actually crossing the horizon itself. The existence of these surfaces suggest that information about the quantum structure of the black hole interior can be inferred through entanglement measures that do not require direct access to the black hole or the cosmological horizon. This points to the possibility that quantum extremal surfaces can serve as indirect probes of regions beyond the horizon, offering a novel perspective on the information encoded in such regions without the need for conventional horizon crossing \cite{Hartman:2013qma,Rozali:2019day,Cooper:2018cmb}. We also obtain the time evolution of the reflected entropy and mutual information from the black hole interior in the presence of the EOW brane, and investigate an indicator of tripartite entanglement termed as the Markov gap \cite{Hayden:2021gno} in this setting. 

	The rest of the manuscript is organized as follows. In \cref{sec:review} we briefly review the DES prescription and the island formula for reflected entropy and provide necessary details of the toy model of braneworld cosmology under consideration. This section also serves to establish the notation used throughout the paper and provides a brief overview of computing entanglement entropies in the presence of EOW branes. In \cref{sec:SR}, we provide a detailed analysis of the reflected entropy for adjacent and disjoint subsystems in the asymptotic boundary from both the DES and island prescriptions and verify their equivalence in the braneworld cosmology. Furthermore in \cref{sec:Plot}, we discuss the time evolution of the reflected entropy and plot the mutual information and the Markov gap. Finally, in \cref{sec:conc}, we summarize our results and comment on possible future directions.

	\section{Review}\label{sec:review}
	
	\subsection{AdS/BCFT and defect extremal surface}
	We begin with a brief review of the salient features of the AdS/BCFT correspondence, first proposed in \cite{Takayanagi:2011zk,Fujita:2011fp}. A boundary conformal field theory (BCFT) is a conformal field theory defined on a manifold $\cM$ with boundary $\partial\cM$ on which conformal boundary conditions are imposed. According to the proposal in \cite{Cardy:2004hm,Cardy:1989ir}, the bulk dual geometry consists of an asymptotically AdS spacetime $\cN$ truncated by a constant tension end-of-world (EOW) brane $\cQ$. In $(d+1)$ spacetime dimensions, the Euclidean action consists of the usual Einstein-Hilbert term on $\cN$ and a Gibbons-Hawking-York term on $\cQ$, along with a worldvolume action \cite{Takayanagi:2011zk,Fujita:2011fp}
	\begin{align}
		I_E=-\frac{1}{16\pi G_N}\int_\cN\sqrt{g}(R-2\Lambda)-\frac{1}{8\pi G_N}\int_\cQ\sqrt{h}(K-(d-1)T)\,,
	\end{align}
	where $K$ denotes the trace of the extrinsic curvature on the EOW brane with tension $T$. The brane trajectory in the bulk geometry may be obtained from varying the above action with respect to $h_{ab}$, the induced metric on the brane. This leads to the Neumann boundary conditions
	\begin{align}
		K_{ab}-Kh_{ab}=-Th_{ab}\,.\label{NBC}
	\end{align}
	Recently, in \cite{Deng:2020ent,Chu:2021gdb}, the AdS/BCFT framework was extended through the inclusion of conformal matter on a tensionless EOW brane, thereby turning on a finite tension. The Neumann boundary condition \eqref{NBC} is modified through the expectation value of the stress tensor of this conformal matter theory and the EOW brane is essentially treated as a defect in the bulk geometry. In the modified bulk geometry, the holographic entanglement entropy of a subsystem $A$ in the BCFT now involves contributions from the defect matter and the usual RT prescription is modified to the defect extremal surface prescription \cite{Deng:2020ent}
	\begin{align}
		S_A=\textrm{min}~\underset{\Gamma,X}{\textrm{ext}}\left[\frac{\textrm{Area}(\Gamma)}{4G_N}+S_\textrm{defect}(D)\right]~~,~~X=\Gamma\cap D\,,\label{DES-EE-formula}
	\end{align}
	where $\Gamma$ is a codimension-2 surface homologous to $A$, and $D$ is the defect on the EOW brane sought out by the bulk entanglement wedge. The authors in \cite{Deng:2020ent,Chu:2021gdb} further demonstrated that the DES prescription leads to the correct entanglement entropy as predicted from the island prescription in the lower dimensional effective description obtained from a partial dimensional reduction.
	\subsection{Defect extremal surface for reflected entropy}
	In this subsection, we review the defect extremal surface prescription as a doubly holographic counterpart of the island formula for the reflected entropy. Recall that, in the presence of an quantum extremal island $\textrm{Is}(A\cup B)$ corresponding to a bipartite state $\rho_{AB}$ in the effective braneworld description, the reflected entropy receives contribution from a quantum extremal cross section $\Gamma=\partial I_{S_R}(A)\cap\partial I_{S_R}(B)$ as follows \cite{Li:2020ceg, Chandrasekaran:2020qtn}
	\begin{figure}[ht]
		\centering
		\includegraphics[scale=0.5]{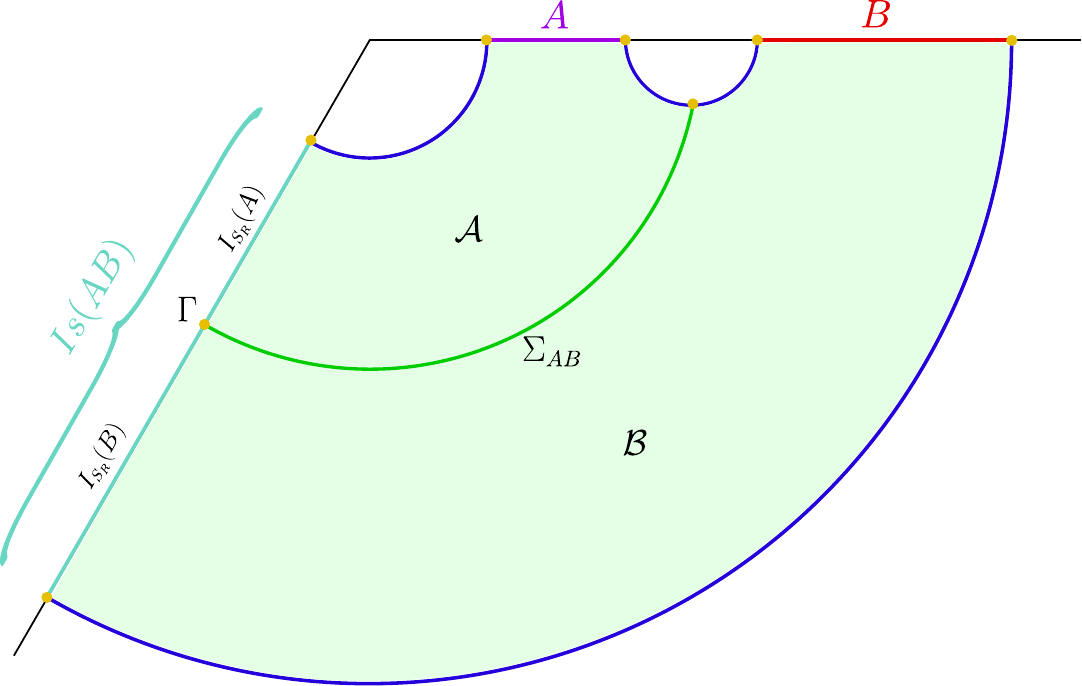}
		\caption{ A schematic representation the island and DES formula for the reflected entropy.
			The green-shaded region represents the entanglement wedge of $A \cup B$, while the green curve is the EWCS which divides the bulk into regions $\mathcal{A}$ and $\mathcal{B}$. The intersection point of the EOW brane and the EWCS is denoted by $\Gamma$. Figure modified from \cite{Li:2021dmf}.} 
		\label{fig:DES}
	\end{figure}
	\begin{equation}\label{SR-bdy}
		S^\textrm{bdy}_{R}(A:B)= \text{min}~ \underset{\Gamma}{\text{ext}} \Bigg[\frac{\text{Area}[\Gamma]}{2 G_N}+S^{\text{eff}}_{R}(A\cup I_{S_{R}}(A):B \cup I_{S_{R}}(B))\Bigg],
	\end{equation}
	where the reflected entropy islands $I_{S_{R}}(A)$ and $I_{S_{R}}(B)$ divide the entanglement island $\textrm{Is}(A\cup B)$ into two parts at the QECS $\Gamma$ (as depicted in \cref{fig:DES}),
	\begin{align}
		\textrm{Is}(A\cup B)=I_{S_{R}}(A)\cup I_{S_{R}}(B)\,.
	\end{align}

	In the doubly holographic perspective, described by the AdS/BCFT setup modified with defect conformal matter on the EOW brane \cite{Deng:2020ent}, the reflected entropy is obtained from the defect extremal surface prescription as \cite{Li:2021dmf}
	\begin{equation}\label{SR-bulk}
		S^\textrm{bulk}_{R}(\mathcal{A}:\mathcal{B})= \text{min}~ \underset{\Sigma_{AB}}{\text{ext}} \Bigg[\frac{\text{Area}[\Sigma_{AB}]}{2 G_N}+S^{\text{eff}}_{R}(\mathcal{A}:\mathcal{B})\Bigg].
	\end{equation}
	where the defect extremal surface $\Sigma_{AB}$ splits the entanglement wedge of $A \cup B$ into two parts $\mathcal{A}$ and $\mathcal{B}$ in the bulk, as depicted in \cref{fig:DES}.
	In the above expression, since the bulk conformal matter is only located on the EOW brane, the effective reflected entropy $S^{\text{eff}}_{R}(\mathcal{A}:\mathcal{B})$ between the bulk quantum matter reduces to that between the reflected entropy islands $I_{S_{R}}(A)$ and $I_{S_{R}}(B)$ on the brane.
	\subsection{Cosmology on the EOW brane}
	Consider a holographic BCFT$_2$ defined on an interval of Euclidean time $[-\tau_0,\tau_0]\times S^1$ for which the boundary states $\ket{B}_\pm$ are called Cardy states. The holographic dual spacetime with the EOW brane corresponds to two saddle geometries: BTZ black hole with connected EOW brane and the thermal AdS geometry for which we have two disconnected EOW branes \cite{Fujita:2011fp,Cooper:2018cmb}. In this note, we consider the BTZ geometry corresponding to high temperatures, whose metric in the Euclidean signature is given as
	\begin{align}
		\d s^2=\frac{r^2-r_h^2}{\ell^2}\d\tau^2+\frac{\ell^2}{r^2-r_h^2}\d r^2+r^2\d\phi^2\,,\label{BTZ-metric}
	\end{align}
	where $\ell$ is the AdS$_3$ radius and the horizon radius $r_h$ is related to the inverse temperature of the black hole $\beta$ as 
	\begin{align}
		r_h=\frac{2\pi\ell^2}{\beta}\,.
	\end{align}
	The EOW brane attached at $\tau=\pm\tau_0$ should preserve the spherical symmetry, which leads to the ansatz $r=r(\tau)$. The Neumann boundary conditions \eqref{NBC} then lead to the following brane trajectory\footnote{We consider solutions which are symmetric about $\tau=0$, namely $\Dot{r}(0)=0$, where the overdot denotes derivative with respect to $\tau$.} \cite{Fujita:2011fp,Wang:2021xih}
	\begin{align}
		r(\tau)=\frac{r_h}{\sqrt{1-T^2}}\sqrt{1+T^2\ell^2\tanh^2\frac{r_h\tau}{\ell^2}}=r_0\sec\frac{r_h\tau^\prime}{\ell^2}\,,\label{brane-trajectory}
	\end{align}
	where $\tau^\prime$ is the brane conformal time\footnote{This may be easily seen from the induced metric on the EOW brane \cite{Wang:2021xih}
		\begin{align}
			\d s^2_\textrm{brane}=\frac{r_0^2}{\cos^2\frac{r_h\tau^\prime}{\ell^2}}\left(\frac{1}{\ell^2}\d\tau^{\prime\,2}+\d\phi^2\right)\,.\label{brane-metric}
	\end{align}} defined through
	\begin{align}
		\tan\frac{r_h\tau^\prime}{\ell^2}=T\ell\tan\frac{r_h\tau}{\ell^2},
	\end{align}
	and we have used the shorthand notation $r_0=\frac{r_h}{\sqrt{1-T^2\ell^2}}$. As described in \cite{Cooper:2018cmb,Wang:2021xih}, the brane trajectory always meets the asymptotic boundary at antipodal points since $\tau_0=\frac{\beta}{4}$.
	
	From symmetry, one may take the $\tau=0$ slice as the initial slice for Lorentzian evolution, taking the Wick rotation $t=-i\tau$. Then the induced metric on the brane may be seen to follow (a $2d$ version of) FRW cosmology \cite{Cooper:2018cmb,Wang:2021xih}
	\begin{align}
		\d s^2_\mathrm{brane}=-\d\lambda^2+r^2(\lambda)\d\phi^2,\label{Brane-cosmology-metric}
	\end{align}
	with the scale factor $r(\lambda)=r_0\cos\frac{r_h\lambda}{\ell r_0}$ describing a big-bang big-crunch cosmology. Note that in writing \eqref{Brane-cosmology-metric}, the rescaled time coordinate $\lambda$ is defined as $\d\lambda=\frac{r^2-r_h^2}{\ell^2T r}\d t$.
	
	\subsubsection{Partial dimensional reduction: effective braneworld cosmology}
\begin{figure}[ht]
		\centering
		\includegraphics[scale=.55]{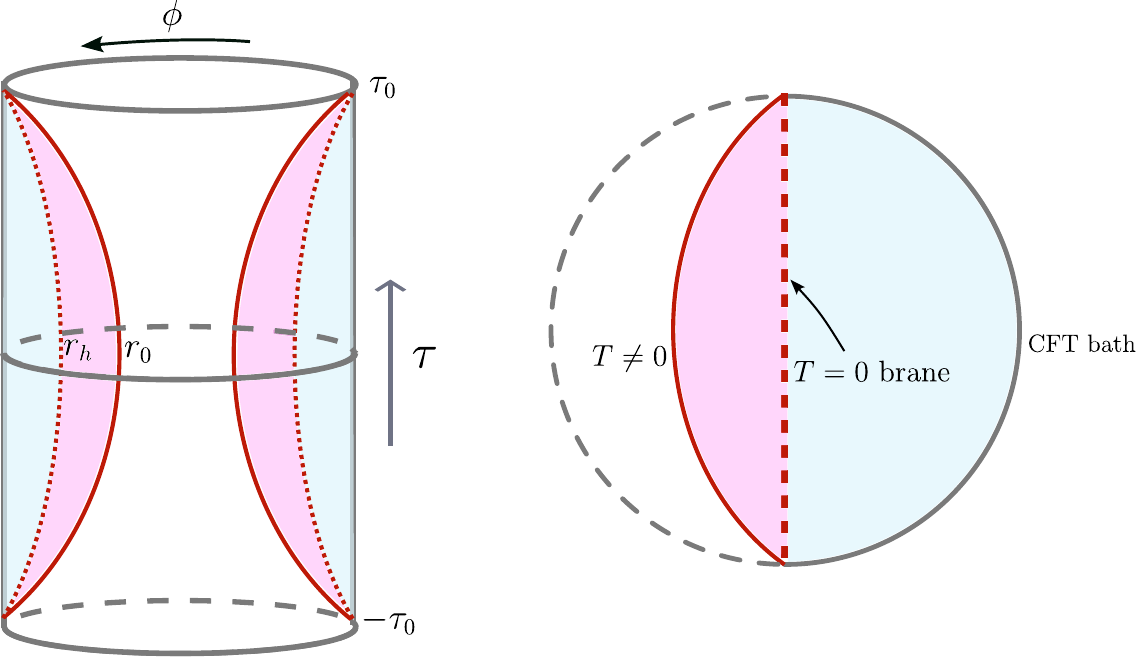}
		\caption{Left: Illustration of the Euclidean geometry describing the EOW brane trajectory. The solid red curve denotes the EOW brane with tension $T$, while the dashed red curve is the zero tension brane. Right: A schematic depiction of the partial reduction of a time slice of AdS$_3$. Figures modified from \cite{Wang:2021xih}.}
		\label{fig:reduction}
	\end{figure}

	In reference \cite{Wang:2021xih}, a lower-dimensional effective theory consisting of the braneworld cosmology coupled to a CFT was obtained through a combination of Randall-Sundrum reduction and AdS/BCFT correspondence. Here, we briefly review their construction. Introducing the zero tension brane
	\begin{align}
		r(\tau^\prime)=r_h\sec\frac{r_h\tau^\prime}{\ell^2},
	\end{align}
	as a transparent interface\footnote{Note that there is no physical degree of freedom on this $T=0$ brane.}, the bulk spacetime is decomposed into two parts:
	\begin{enumerate}
		\item The reduction region shown as shaded red colour in \cref{fig:reduction}, where one employs partial Randall-Sundrum reduction to obtain an effective gravitational theory on the EOW brane along with a defect matter theory described by the same CFT as that on the asymptotic boundary.
		\item The dual region depicted as shaded light blue colour in \cref{fig:reduction} , where one employs standard AdS/BCFT technique to obtain a BCFT with zero boundary entropy corresponding to the zero tension brane.
	\end{enumerate}
	These two theories, defined on a hybrid manifold comprising a fluctuating metric on the gravitating background and a flat CFT bath, are naturally coupled at the interface through transparent boundary conditions. This procedure is illustrated in \cref{fig:reduction}.
	
	In particular, an effective Newton's constant on the braneworld gravity may be obtained by integrating the Warp factor from the zero tension brane to the finite tension brane \cite{Wang:2021xih}. To perform this calculation, it is convenient to transform to the Kruskal-like coordinates $(s,y)$, given in \cref{AppA}, in which the maximally extended black hole spacetime has the form 
	\begin{align}
		\d s^2=\frac{1}{\cos^2 y}\left(-\ell^2\d s^2+\ell^2\d y^2+r_h^2\cos^2s\,\d\phi^2\right)\,,\label{BTZ-Kruskal}
	\end{align}
	in which the EOW brane resides on a constant $y$ slice
	\begin{align}
		y=-\arcsin(T\ell)\,.
	\end{align}
	One may now obtain
	\begin{align}
		\frac{1}{4G_\mathrm{brane}}=\frac{\ell}{4G_N^{(3)}}\int_{-\arcsin(T\ell)}^0\frac{\d y}{\cos y}=\frac{\ell}{4G_N^{(3)}}\log\sqrt{\frac{1+T\ell}{1-T\ell}}\,.\label{G-brane}
	\end{align}
	\subsubsection{Entanglement entropy}\label{subsec;review EE}
	In this subsection, following \cite{Wang:2021xih} we review the computation of entanglement entropy for a interval $A=\left[(\phi_1,\frac{\tau_1}{\ell}),(\phi_2,\frac{\tau_1}{\ell})\right]$ on a fixed time slice $\tau=\tau_1$ of the asymptotic boundary of the braneworld cosmology discussed in the previous subsection.
	\subsubsection*{Bulk defect extremal surface}
	For a large interval, the defect conformal matter 
	on the EOW brane contributes to the entanglement entropy. The candidate defect extremal surface $\Gamma$ ends on the EOW brane and seeks out a portion $D=\left[\left(\phi_1,\frac{\tau_1^\prime}{\ell}\right),\left(\phi_2,\frac{\tau_1^\prime}{\ell}\right)\right]$ of the defect. The contribution of this defect to the entanglement entropy $S_A$ is given, in terms of a two-point function of twist operators inserted at the endpoints of $D$, by
	\begin{align}
		S_D=\lim_{n\to 1}\frac{1}{1-n}\log\left<\sigma_n(\phi_1,\tau_1^\prime)\bar{\sigma}(\phi_2,\tau_1^\prime)\right>_{\textrm{BCFT}^{\otimes n}}
	\end{align}
	In \cite{}, the contribution from the defect matter was shown to be a constant irrespective of the subsystem geometry, which we review below. The induced metric on the brane is conformal to that of a cylinder:
	\begin{align}
		\d s^2_\textrm{brane}=\Omega^{-2}(\tau^\prime)\d s^2_\textrm{cylinder}~~,~~\Omega(\tau^\prime)=\frac{1}{r_0}\cos\left(\frac{r_H\tau^\prime}{\ell^2}\right)\,,\label{brane-conformal-metric}
	\end{align}
	where 
	\begin{align}
		\d s^2_\textrm{cylinder}=\frac{1}{\ell^2}\d\tau^{\prime\,2}+\d\phi^2=\d w^\prime\d\bar{w}^\prime
	\end{align}
	with $w=\phi+i\frac{\tau^\prime}{\ell}$. This cylinder is subsequently mapped to the upper-half-plane (UHP) utilizing the conformal map
	\begin{align}
		w^\prime=-i\frac{\tau_0}{\ell}+\frac{\ell}{r_h}\log z\,,\label{UHP-map}
	\end{align}
	leading to
	\begin{align}
		\d s^2_\textrm{cylinder}=\frac{\ell^2}{r_h^2|z|^2}\d z\d\bar{z}\,.\label{UHP-metric}
	\end{align}
	From \cref{brane-conformal-metric,UHP-metric}, one may identify the total conformal factor relative to the UHP metric as
	\begin{align}
		\hat{\Omega}(z)\Omega(\tau^\prime)=\left|\frac{r_H z}{\ell}\frac{1}{r_0}\cos\left(\frac{r_h\tau^\prime}{\ell^2}\right)\right|\,.\label{Conformal-factors}
	\end{align}
	The two-point function of twist operators on the UHP is translated into a four-point function of chiral twist operators, which may be subsequently expanded in terms of bulk or boundary intermediate operators. In the large central charge limit, assuming vacuum block dominance, the UHP two-point function then has the following form \cite{Sully:2020pza} 
	\begin{equation}\label{OPE-BOE}
		\left<\sigma_n(z_1,\bar{z}_1)\bar{\sigma}_n(z_2,\bar{z}_2)\right>_{\textrm{UHP}^{\otimes n}}=\begin{cases}
			\left(\frac{|z_1-z_2|}{\epsilon_b}\right)^{-2d_n}~~~&\textrm{bulk channel}\notag\\
			g_b^{2(1-n)}\left(\frac{4\,\textrm{Im}z_1\textrm{Im}z_2}{\epsilon_b^2}\right)^{-d_n}~~~&\textrm{boundary channel}
		\end{cases}
	\end{equation}
	where $\epsilon_b$ is the UV cut-off on the EOW brane and $d_n=\frac{c}{12}\left(n-\frac{1}{n}\right)$ is the conformal dimension of the twist operators.
	
	When the defect theory contributes, the entanglement wedge must contain the defect, and hence the boundary channel is favored. Utilizing \cref{OPE-BOE} and the conformal factors \eqref{Conformal-factors}, we may obtain the defect contribution to be
	\begin{align}
		S_D=\frac{c}{3}\log\left(\frac{2r_0\ell}{r_h\epsilon_b}\right)=\frac{c}{3}\log\left(\frac{2\ell}{\epsilon_b\sqrt{1-T^2 \ell^2}}\right),\,\label{defect-contribution}
	\end{align}
	a constant, as advertised earlier. Hence, according to \cref{DES-EE-formula}, the defect extremal surface is solely given by the extremal geodesics emanating from the endpoints of the subsystem $A$ and landing on the EOW brane. 
	
	To obtain the length of these geodesics, as well as to facilitate later computations, recall that any asymptotically AdS geometry can be embedded in $\mathbb{R}^{2,2}$: 
	\begin{align}
		\d s^2=\eta_{AB}\d x^A\d X^B=-\left(\d X^0\right)^2-\left(\d X^1\right)^2+\left(\d X^2\right)^2+\left(\d X^3\right)^2,\label{embedding-coordinates}
	\end{align}
	subject to the quadratic constraint $X^2=-\ell^2$. In particular, for the BTZ black hole with metric given in \cref{BTZ-metric}, the embedding coordinates are given by (we have shifted to the Lorentzian signature)
	\begin{align}\label{embedding-r}
		X^0&=\ell\sqrt{\frac{r^2}{r_h^2}-1}\sinh\left(\frac{r_h t}{\ell^2}\right)\,,\notag\\
		X^1&=\ell\frac{r}{r_h}\cosh\left(\frac{r_h \phi}{\ell}\right)\,,\notag\\
		X^2&=\ell\frac{r}{r_h}\sinh\left(\frac{r_h \phi}{\ell}\right)\,,\notag\\
		X^3&=\ell\sqrt{\frac{r^2}{r_h^2}-1}\cosh\left(\frac{r_h t}{\ell^2}\right)\,.
	\end{align}
	The length of any geodesic between two bulk points $\left(\phi_1,\frac{t_1}{\ell},r_1\right)$ and $\left(\phi_2,\frac{t_2}{\ell},r_2\right)$ may be obtained as
	\begin{align}
		\mathcal{L}_{12}=\ell\,\textrm{arccosh} \left(\zeta_{12}\right)\,,\label{BTZ-geodesic-length}
	\end{align}
	where $\zeta_{12}$ is the unique invariant associated with the two bulk points,
	\begin{align}
		\zeta_{12}&=-\frac{1}{\ell^2}X\left[\phi_1,t_1,r_1\right]\cdot X\left[\phi_2,t_2,r_2\right]\notag\\\
		&=\frac{r_1 r_2}{r_h^2} \cosh \left(\frac{r_h \left(\phi_1-\phi_2\right)}{\ell}\right)-\sqrt{\left(\frac{r_1^2}{r_h^2}-1\right)\left(\frac{r_2^2}{r_h^2}-1\right)}\cosh \left(\frac{r_h\left(t_1-t_2\right)}{\ell^2}\right)\,.\label{BTZ-invariant}
	\end{align}
	
	Now we consider the geodesic between the boundary point $\left(\phi_1,\frac{t_1}{\ell},\frac{\ell^2}{\epsilon}\right)$ and an arbitrary point $\left(\phi_b,\frac{t_b}{\ell},r_b\right)$ on the EOW brane, where $r_b$ and $t_b$ satisfies the constraint \eqref{brane-trajectory}. The length of this geodesic may be easily obtained from \cref{BTZ-geodesic-length,BTZ-invariant} as
	\begin{align}
		\cL_\Gamma=\ell\,\log\left[2\ell^2\frac{\sqrt{1-\ell^2 T^2 \tanh ^2\left(\frac{r_h t_b}{l^2}\right)}\cosh \left(\frac{r_h (\phi_1-\phi_b)}{\ell}\right)-T \ell \,\text{sech}\left(\frac{r_h t_b}{\ell^2}\right)\cosh \left(\frac{r_h(t_1-t_b)}{\ell^2}\right)}{r_h^2\sqrt{1 - \ell^2 T^2}\epsilon}\right]\,.
	\end{align}
	Extremizing $\cL_\Gamma$ with respect to the arbitrary parameters $\phi_b,t_b$, we obtain the following solutions
	\begin{align}
		\phi_b=\phi_1~~,~~T\ell\tanh\left(\frac{r_ht_b}{\ell^2}\right)=-\tanh\left(\frac{r_ht_1}{\ell^2}\right).\label{HEE-extremization}
	\end{align}
	Therefore the length of the extremal surface is given by
	\begin{align}\label{Extremal-length}
		\cL_\Gamma^\textrm{ext}=\ell\log\left[\frac{2\ell^2}{\epsilon r_h}\cosh\left(\frac{r_h t_1}{\ell^2}\right)\sqrt{\frac{1+T\ell}{1-T\ell}}\right],
	\end{align}
	which is clearly independent of the location of the boundary endpoint $\phi_1$. Adding the contribution from the geodesic extending from the other boundary endpoint as well as the defect contribution in \cref{defect-contribution}, we may obtain the entanglement entropy of the subsystem $A$ as
	\begin{align}
		S_A&=\frac{\cL_\Gamma^\textrm{ext}}{4G_N^{(3)}}+S_D\notag\\
		&=\frac{c}{3}\log\left[\frac{2\ell^2}{\epsilon r_h}\cosh\left(\frac{r_h t_1}{\ell^2}\right)\sqrt{\frac{1+T\ell}{1-T\ell}}\right]+\frac{c}{3}\log\left(\frac{2\ell}{\epsilon_b\sqrt{1-T^2 \ell^2}}\right)\,.\label{DES-result}
	\end{align}
	
	On the other hand, for a small interval, the entanglement wedge does not include any defect on the EOW brane, and the defect extremal surface reduces to the usual RT surface. The length of this RT surface joining the endpoints of the subsystem may be readily obtained from \cref{BTZ-geodesic-length,BTZ-invariant}. The entanglement entropy is then given as
	\begin{align}
		S_A=\frac{c}{3}\log\left[\frac{2\ell^2}{\epsilon r_h}\sinh\left(\frac{r_h(\phi_2-\phi_1)}{2\ell}\right)\right]\,.\label{No-defect-result}
	\end{align}
	\subsubsection*{Islands in cosmology}
	In the effective braneworld theory comprised of a cosmological spacetime coupled with a non-gravitating thermal bath, we may use the island prescription to compute the entanglement entropy for the subsystem $A$ in the thermal bath as follows \cite{Almheiri:2019qdq,Almheiri:2019yqk,Penington:2019npb,Penington:2019kki}
	\begin{align}
		S_A=\textrm{min}~\underset{I}{\textrm{ext}}\left[\frac{\textrm{Area}(\partial I)}{4G_\textrm{brane}}+S_\textrm{eff}(A\cup I)\right]\,.
	\end{align}
	As discussed in \cite{Wang:2021afl}, in the effective theory the brane conformal time $\tau^\prime$ may be treated as a natural extension of the bath time $\tau$ and the island may be chosen as $I=[w_1^I,w_2^I]=\left[\left(\phi_1,\frac{\tau_0+\tau_I}{\ell}\right),\left(\phi_2,\frac{\tau_0+\tau_I}{\ell}\right)\right]$. Accounting for the two endpoints of the island, the area term in the above expression may be computed from \cref{G-brane} as follows
	\begin{align}\label{Area-term}
		\frac{\textrm{Area}(\partial I)}{4G_\textrm{brane}}=2\times\frac{1}{G_\textrm{brane}}=\frac{c}{3}\log\sqrt{\frac{1+T\ell}{1-T\ell}}.
	\end{align}
	On the other hand, in the large central charge limit, the 4-point twist correlator computing the effective matter entropy may be factorized into two 2-point functions of twist operators, leading to 
	\begin{align}
		S_\textrm{eff}(A\cup I)=\lim_{n\to 1}\frac{1}{1-n}\log\left[\left<\sigma_n(w_1,\bar{w}_1)\bar\sigma_n(w^I_1,\bar{w}^I_1)\right>_{\textrm{CFT}^{\otimes n}}\left<\sigma_n(w_2,\bar{w}_2)\bar\sigma_n(w^I_2,\bar{w}^I_2)\right>_{\textrm{CFT}^{\otimes n}}\right].
	\end{align}
	As earlier, the brane CFT may be mapped to the UHP using the series of conformal maps given in \cref{brane-conformal-metric,UHP-map}. Furthermore, the bath CFT could be mapped to the down-half plane (DHP) using another conformal transformation
	\begin{align}\label{CFT_DHP}
		w=i\frac{\tau_0}{\ell}+\frac{\ell}{r_h}\log z\,,
	\end{align}
	which is tantamount to a conformal factor 
	\begin{align}\label{Conformal-factor-bath-cft}
		\Tilde{\Omega}\cdot \hat{\Omega}(z)=\frac{1}{\ell}\left|\frac{r_h z}{\ell}\right|\,.
	\end{align}
	utilizing all these transformations, the hybrid manifold is transformed to the complex plane and the correlation functions of the twist operators are readily computed. Extremizing the generalized entropy leads to the solution
	\begin{align}\label{brane-time-EE}
		\tau_I=\frac{\pi\ell^2}{2r_h}-\tau_1\,,
	\end{align}
	and the entanglement entropy is given by
	\begin{align}
		S_A=\frac{c}{3}\log\sqrt{\frac{1+T\ell}{1-T\ell}}+\frac{c}{3}\log\left[\frac{2\ell^2}{r_h\epsilon}\cos\left(\frac{r_h\tau_1}{\ell^2}\right)\right]+\frac{c}{3}\log\left(\frac{2r_0\ell}{r_h\epsilon_b}\right).
	\end{align}
	Upon employing the Wick rotation $t_1=-i\tau_1$, the above expression matches identically with \eqref{DES-result} obtained from the bulk DES formula, thereby validating the proposal \eqref{DES-EE-formula}. 
	
	On the other hand, for a small interval incapable of accommodating an island on the brane, the entanglement entropy is computed from the effective matter entropy which conforms with the bulk computations in \cref{No-defect-result}.
	
	\section{Holographic reflected entropy }\label{sec:SR}
	
	\subsection{Adjacent Subsystems}\label{sec:adjacent}
	In this subsection, we examine the reflected entropy for two adjacent subsystem
	$A=\left[(\phi_1,\frac{\tau_1}{\ell}),(\phi_2,\frac{\tau_1}{\ell})\right]$ and $B=\left[(\phi_2,\frac{\tau_1}{\ell}),(\phi_3,\frac{\tau_1}{\ell})\right]$ on a time slice $\tau=\tau_1$ of the asymptotic boundary of the braneworld cosmology discussed in \cref{sec:review}. We observe that for two adjacent subsystems there are two possible phases of the EE depending on the subsystems size.
	In the following, we provide detailed computations of the various reflected entropy phases, obtained from both the boundary and bulk perspectives.
	
	The bulk computation of the minimal EWCS is greatly simplified in asymptotically AdS$_3$ spacetimes, as minimal surfaces are given by geodesics. In the following, we will heavily make use of the minimal geodesic distance between a boundary anchored geodesic and an arbitrary bulk point in asymptotically AdS$_3$ spacetimes. This is most efficiently computed in the embedding coordinate formalism, wherein the minimal geodesic distance between a geodesic connecting two points $X_1^A$ and $X_2^A$ on the asymptotic boundary, and an arbitrary bulk point $X_2^A$ is computed through the expression \cite{Basu:2023jtf}
	\begin{align}\label{EWCS-Adj}
		\mathcal{L}\left(X_2,X_{13}\right)=\cosh^{-1} \left(\sqrt{\frac{2 \zeta_{12} \zeta_{23}}{\zeta_{13}}}\right),
	\end{align}
	where $\zeta_{ij}=-X_i\cdot X_j$.

	\subsubsection{Entanglement entropy phase 1}\label{EE-adj1}
	\begin{figure}[ht]
		\centering
		\includegraphics[scale=1.2]{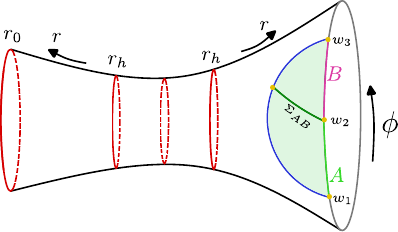}
		\caption{Diagrammatic illustration of the EE phase $1$ in a time slice $t=t_1$, when the RT surface and the EWCS for
			$A \cup B$ are shown as blue and green curves. The connected entanglement wedge is depicted by the green shaded region. } 
		\label{fig:adj1}
	\end{figure}
	We begin with the case when both the subsystems are small and close to each other. Hence the EE for this phase is proportional to the length of a dome-shaped RT surface shown by the blue curve in \cref{fig:adj1}. 
	Substituting the end points $w_1=(\phi_{1},\frac{\tau_1}{\ell})$ and $w_3=(\phi_{3},\frac{\tau_1}{\ell})$ of the blue curve in 
	\cref{BTZ-invariant}, the EE for this phase may be obtained as follows
	\begin{align}
		S^{(1)}_{AB}= \frac{1}{2 G_N}\log \left[\frac{2 \ell^2}{r_h \epsilon}\sinh \left(\frac{r_h \phi_{31}}{2\ell}\right)\right] .
	\end{align}
	For this case there is only one reflected entropy or the EWCS phase, depicted by the green curve in \cref{fig:adj1}. In the boundary perspective, the reflected entropy may be computed by using three-point twist field correlator as follows\footnote{In the following, we will use the shorthand $mn$ or $m$ to signify the fact that the correlation functions are evaluated on the orbifold theories $\textrm{CFT}^{\bigotimes mn}$ or $\textrm{CFT}^{\bigotimes m}$.} \cite{Dutta:2019gen} 
	\begin{align}
		S_R^{\text{bdy}}(A:B)= \lim_{{m,n} \to 1}\frac{1}{1-n}\log\frac{\langle \sigma_{g_A}(w_1, \bar{w}_1)\sigma_{g_A^{-1}g^{}_{B}}(w_2,\bar{w}_2)\sigma_{g_B}(w_3,\bar{w}_3)\rangle_{mn}}{\left(\langle \sigma_{g_m}(w_1,\bar{w}_1)\sigma_{g_m}(w_3,\bar{w}_3)\rangle_{m}\right)^{n}}.
	\end{align}
	Since the BCFT$_2$ is defined on a circle, therefore the computation of the above twist field correlator is not straightforward. To compute the above expression first we need to transform this to the complex plane twist field correlator which may be done by using \cref{CFT_DHP}. Hence the above expression may be written as the complex plane twist field correlator with appropriate conformal factor as follows 
	\begin{align}
		S_R^{\text{bdy}}(A:B)= \lim_{{m,n} \to 1}\frac{1}{1-n}\log\frac{ \left(\tilde{\Omega} ~ \hat{\Omega}(z_2)\right)^{2 h_{AB}}  \langle \sigma_{g_A}(z_1, \bar{z}_1)\sigma_{g_A^{-1}g^{}_{B}}(z_2,\bar{z}_2)\sigma_{g_B}(z_3,\bar{z}_3)\rangle_{mn}}{\left(\langle \sigma_{g_m}(z_1,\bar{z}_1)\sigma_{g^{-1}_m}(z_3,\bar{z}_3)\rangle_m\right)^{n}},
	\end{align}
	where $\tilde{\Omega}=\frac{1}{\ell}$ and $\hat{\Omega}(z_2)=\left|\frac{r^{}_h z_2}{\ell}\right|$ are the conformal factors corresponding to the composite twist operator $\sigma_{g^{-1}_Ag^{}_B }(z_2)$, while other conformal factors corresponding to the points $z_{1}$ and $z_{3}$ cancel from the denominator.
	It is well known that the three-point function on the complex plane has the following form \cite{Dutta:2019gen}
	\begin{align}
		\langle \sigma_{g_A}(z_1, \bar{z}_1)\sigma_{g_A^{-1}g^{}_{B}}(z_2,\bar{z}_2)\sigma_{g_B}(z_3,\bar{z}_3)\rangle_{mn}=  C_{n, m}\left(z_{12} \bar{z}_{12}\right)^{-h_{AB}} \left(z_{23} \bar{z}_{23}\right)^{-h_{AB}}\left(z_{13} \bar{z}_{13}\right)^{2h-h_{AB}},
	\end{align}
	where $z_{ij}= z_i-z_j\,,\,\bar{z}_{ij}= \bar{z}_i-\bar{z}_j$,
	and $h$, $h_{AB}$ are the conformal dimensions of the twist operators $\sigma_{g^{}_A}\,,\,\sigma_{g^{}_B}$ and the composite operator $\sigma_{g_A^{-1}g_{B}}$ respectively.
	The conformal dimensions of the twist operators and the OPE coefficient $C_{n,m}$ are given by \cite{Dutta:2019gen}
	\begin{align}\label{conformal dimensions}
		h=\frac{n c}{24}\left(m-\frac{1}{m}\right)\equiv n h_{m}~~,~~ h_{AB}=\frac{2c}{24}\left(n-\frac{1}{n}\right)~~,~~C_{n,m}\equiv C_{\sigma_{g^{}_A}\sigma_{g^{-1}_Ag^{}_B }\sigma_{g^{}_B}}=(2m)^{-4h}.
	\end{align}
	Now using the form of the three-point and two-point twist field correlators and restoring the original coordinates, the reflected entropy between the adjacent subsystems in this phase may be obtained as
	\begin{align}\label{SR_Adj1}
		S^{\textrm{bdy}}_R(A:B)= \frac{c}{3}\log \left[\frac{4\ell^2}{r_h \epsilon}\,\frac{\sinh \left(\frac{r_h \phi_{21}}{2 \ell}\right) \sinh \left(\frac{r_h \phi_{32}}{2 \ell}\right)}{\sinh\left(\frac{r_h \phi_{31}}{2 \ell}\right)}\right].
	\end{align}
	In the bulk perspective, the EWCS is proportional to the length of the green geodesic $\Sigma_{AB}$, shown in \cref{fig:adj1}. Now by utilizing the embedding coordinates given in \cref{embedding-r} corresponding to the end-points of the subsystems in \cref{EWCS-Adj}, the bulk EWCS in this phase may be obtained as   
	\begin{align}
		S^{\text{bulk}}_{R}(\mathcal{A}:\mathcal{B})= \frac{1}{2 G_N} \log \left[\frac{2 \ell^2 \left(e^{\frac{r_h \phi_{21}}{\ell}}-1\right) \left(e^{\frac{r_h \phi_{32}}{\ell}}-1\right)}{r_h \epsilon  \left(e^{\frac{r_h \phi_{31}}{\ell}}-1\right)}\right].
	\end{align}
	Note that upon utilizing the Brown-Henneaux relation, the above expression exactly matches with the reflected entropy computed in \cref{SR_Adj1}.

	\subsubsection{Entanglement entropy phase 2}\label{EE-adj2}
	\begin{figure}[ht]
		\centering
		\includegraphics[scale=1.2]{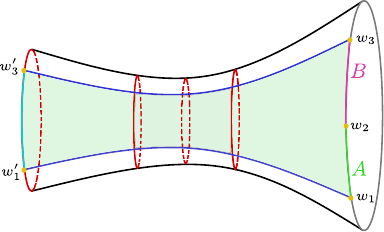}
		\caption{Schematic illustration of the EE phase 2 when the RT surface for $A \cup B$ are shown as blue curves and the entanglement wedge is the shaded green region.} 
		\label{fig:adj2}
	\end{figure}

	In this phase, we assume that both the subsystems are large and far away from each other, hence the EE is given by the length of two RT surfaces which crosses the horizon and end at the EOW brane, shown as solid blue curve in \cref{fig:adj2}. Now using \cref{DES-result}, the EE for this phase may be written as
	\begin{align}
		S^{(2)}_{AB}=\frac{1}{2 G_N}\left(\log \left[\frac{2\ell^2}{r_h\epsilon}\cosh \left(\frac{r_h t_1}{\ell^2}\right)\right]+\log \left(\frac{2 \ell}{\epsilon_b\sqrt{1- T^2 \ell^2}}\right)+\log \sqrt{\frac{1+ T \ell }{1-T \ell }}\right).
	\end{align}
	In this EE phase we observe three possible phases of the reflected entropy or the bulk EWCS, depending on the size of the subsystems. In the following, we detail the computations of the reflected entropy for each phase from both the boundary and bulk perspectives.

	\subsubsection*{Phase-I}\label{SR_adj_EE2(i)}
\begin{figure}[H]
		\centering
		\includegraphics[scale=1.2]{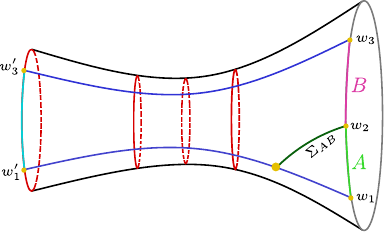}
		\caption{ Schematic illustration of the bulk EWCS between subsystems $A$ and $B$, depicted as solid green curve.} 
		\label{fig:adj2(i)}
	\end{figure}

	\paragraph*{The boundary perspective:}  
	For this reflected entropy phase, the subsystem $A$ is much smaller compared to subsystem $B$, and therefore the EWCS lands on the extremal surface corresponding to the points $w_1$ and $w^{\prime}_1$.
	As seen from \cref{fig:adj2(i)}, there is no island cross section for this case, hence the reflected entropy in the boundary description may be obtained from the following twist field correlators 
	\begin{align}\label{SR-adj2(i)-factor}
		&S^{\text{(bdy)}}_{R}(A:B)=S^{\text{eff}}_{R}(A:B \cup I_{S_{R}}(B))\notag\\&= \lim_{{m,n} \to 1}\frac{1}{1-n}\log\frac{\langle \sigma_{g_A}(w_1,\bar{w}_1)\sigma_{g^{}_A g^{-1}_B}(w^{}_2,\bar{w}^{}_2)\sigma_{g^{-1}_A}(w^{I}_1,\bar{w}^{I}_1)\sigma_{g^{-1}_B}(w^{}_3,\bar{w}^{}_3)\sigma_{g^{}_B}(w^{I}_3,\bar{w}^{I}_3)\rangle_{mn}}{\langle \sigma_{g_m}(w_1,\bar{w}_1)\sigma_{g^{-1}_m}(w^{I}_1,\bar{w}^{I}_1)\sigma_{g^{-1}_m}(w^{}_3,\bar{w}^{}_3)\sigma_{g^{}_m}(w^{I}_3,\bar{w}^{I}_3)\rangle^{n}_{m}}.
	\end{align}
	Here $w^{I}_1=(\phi_{1},\frac{\tau_0+\tau^{I}_1}{\ell})$ is a point on the EOW brane where the extremal surface corresponding to the endpoint $w_1$ of $A\cup B$ intersects with the EOW brane. Recall that, extremization of the EE for $A\cup B$ requires that $\tau_1^I$ is given by \cref{brane-time-EE}. In the large central charge limit, numerator of the above expression may be factorized into one three-point and one two-point twist field correlator as follows \cite{Chandrasekaran:2020qtn,Li:2020ceg}
	\begin{align}
		&\langle \sigma_{g_A}(w_1,\bar{w}_1)\sigma_{g^{}_A g^{-1}_B}(w^{}_2,\bar{w}^{}_2)\sigma_{g^{-1}_A}(w^{I}_1,\bar{w}^{I}_1)\sigma_{g^{-1}_B}(w^{}_3,\bar{w}^{}_3)\sigma_{g^{}_B}(w^{I}_3,\bar{w}^{I}_3)\rangle_{mn}\notag\\
		& = \langle \sigma_{g_A}(w_1,\bar{w}_1)\sigma_{g^{}_A g^{-1}_B}(w^{}_2,\bar{w}^{}_2)\sigma_{g^{-1}_A}(w^{I}_1,\bar{w}^{I}_1)\rangle_{mn}\langle \sigma_{g^{-1}_B}(w^{}_3,\bar{w}^{}_3)\sigma_{g^{}_B}(w^{I}_3,\bar{w}^{I}_3)\rangle_{mn},
	\end{align}
	while the denominator is factorized into two two-point twist field correlators
	\begin{align}
		&\langle \sigma_{g_m}(w_1,\bar{w}_1)\sigma_{g^{-1}_m}(w^{I}_1,\bar{w}^{I}_1)\sigma_{g^{-1}_m}(w^{}_3,\bar{w}^{}_3)\sigma_{g^{}_m}(w^{I}_3,\bar{w}^{I}_3)\rangle^{n}_{m}\notag\\
		& = \langle \sigma_{g_m}(w_1,\bar{w}_1)\sigma_{g^{-1}_m}(w^{I}_1,\bar{w}^{I}_1)\rangle^{n}_{m}\langle\sigma_{g^{-1}_m}(w^{}_3,\bar{w}^{}_3)\sigma_{g^{}_m}(w^{I}_3,\bar{w}^{I}_3)\rangle^{n}_{m}.
	\end{align} 
	Substituting the above expressions into \cref{SR-adj2(i)-factor}, the reflected entropy for this phase may be computed as
	\begin{align}
		S^{\text{(bdy)}}_{R}(A:B)=\lim_{{m,n} \to 1}\frac{1}{1-n}\log\frac{\langle \sigma_{g_A}(w_1,\bar{w}_1)\sigma_{g^{}_B g_A^{-1}}(w_2,\bar{w}_2)\sigma_{g^{-1}_A}(w^{I}_1,\bar{w}^{I}_1)\rangle_{mn}}{\langle \sigma_{g_m}(w_1,\bar{w}_1)\sigma_{g^{-1}_m}(w^{I}_1,\bar{w}^{I}_1)\rangle^{n}_{m}}.
	\end{align}
	Using the conformal transformations given in \cref{CFT_DHP,UHP-map}, the above expression may be rewritten in terms of a correlator on the complex plane:
	\begin{align}
		S_R^{\text{bdy}}(A:B)= \lim_{{m,n} \to 1}\frac{1}{1-n}\log\frac{ \left(\tilde{\Omega} ~ \hat{\Omega}(z_2)\right)^{2 h_{AB}}  \langle \sigma_{g_A}(z_1, \bar{z}_1)\sigma_{g_{B} g_A^{-1}}(z_2,\bar{z}_2)\sigma_{g_B}(z^{I}_1,\bar{z}^{I}_1)\rangle_{mn}}{\langle \sigma_{g_m}(z_1,\bar{z}_1)\sigma_{g^{-1}_m}(z^{I}_1,\bar{z}^{I}_1)\rangle^{n}_m}.
	\end{align}
	Here $\tilde{\Omega}$ and $\hat{\Omega}(z_2)$ are the conformal factors given in \cref{Conformal-factor-bath-cft}.  Utilizing the form of the three-point twist field correlator and substituting the brane conformal time given in \cref{brane-time-EE} in the resulting expression, the reflected entropy for this phase may be obtained as
	\begin{align}\label{SR_adj_2(i)}
		S^{\text{bdy}}_{R}(A:B)= \frac{c}{3} \log \left[\frac{2 \sqrt{2} \ell^2}{r_h \epsilon}\sec \left(\frac{r_h \tau_1}{\ell^2}\right) \sinh \left(\frac{r_h \phi_{21}}{2 \ell}\right) \sqrt{\cos\left( \frac{2 r_h \tau_1}{\ell^2}\right)+\cosh \left(\frac{r_h \phi_{21}}{\ell}\right)}\right],
	\end{align}

	\paragraph*{The bulk perspective:}

	In the bulk description, the curve $\Sigma_{AB}$ connects the point $w_2$ to an arbitrary point $w_{\text{HM}}$ on the extremal surface corresponding to points $w_1$ and $w^{\prime}_1$. In this case, the first term of \cref{SR-bulk} vanishes since the entire island belongs to the subsystem $B$. As a result the reflected entropy for this configuration in the bulk description is proportional to the minimal length of the of the curve $\Sigma_{AB}$, shown as solid green curve in \cref{fig:adj2(i)}. 
	
	To compute the length of this curve, we first note that the RT surface connecting the boundary endpoint $w_1$ and the point $w_1^\prime$ on the EOW brane is essentially the well known Hartman-Maldacena (HM) surface introduced in \cite{Hartman:2013qma}. To see this, note that the profile of the usual HM surface joining the points $\left(\phi_1,\frac{t_1}{\ell}\right)$ and $\left(\phi_1,\frac{-t_1+\frac{i\beta}{2}}{\ell}\right)$ on the two asymptotic boundaries is given as
	\begin{align}
		\sqrt{1-\frac{r_h^2}{r^2}}=\frac{\sinh\left(\frac{r_ht_1}{\ell^2}\right)}{\sinh\left(\frac{r_ht}{\ell^2}\right)}~~,~~\phi=\phi_1\,.\label{HM-equation}
	\end{align}
	Utilizing the extremal solution \eqref{HEE-extremization}, it is easy to verify that the coordinates of the endpoint $w_1^\prime$ on the EOW brane satisfies 
	\begin{align}
		\left(1-\frac{r_h^2}{r_b^2}\right)\sinh^2\left(\frac{r_h t_b}{\ell^2}\right)=\sinh^2\left(\frac{r_h t_1}{\ell^2}\right)\,,
	\end{align}
	confirming our claim. Next, we choose an arbitrary point $w_\textrm{HM}=\left(\phi_1,\frac{\hat{t}}{\ell},\hat{r}\right)$ on this HM surface. The geodesic distance between $w_2$ and $w_\textrm{HM}$ may be obtained utilizing \cref{BTZ-geodesic-length,BTZ-invariant} as follows
	\begin{align}
		\mathcal{L}\left(w_2,w_\textrm{HM}\right)&=\ell\,\textrm{arccosh}\left[\frac{\ell^2\hat{r}}{\epsilon r_h^2}\cosh\left(\frac{r_h\phi_{21}}{\ell}\right)-\frac{\ell^2\sqrt{\hat{r}^2-r_h^2}}{\epsilon r_h^2}\cosh\left(\frac{r_h(t_1-\hat{t})}{\ell^2}\right)\right]\notag\\
		&=\ell\log\left[\frac{2\ell^2}{r_h\epsilon}\frac{\cosh\left(\frac{r_h\phi_{21}}{\ell}\right)\sinh\left(\frac{r_h\hat{t}}{\ell^2}\right)-\cosh\left(\frac{r_h(t_1-\hat{t})}{\ell^2}\right)\sinh\left(\frac{r_h t_1}{\ell^2}\right)}{\sqrt{\sinh^2\left(\frac{r_h\hat{t}}{\ell^2}\right)-\sinh^2\left(\frac{r_h t_1}{\ell^2}\right)}}\right]
	\end{align}
	Extremizing the above length with respect to the remaining parameter $\hat{t}$, we obtain the extremal solution to be
	\begin{align}
		\hat{t}=\frac{\ell^2}{r_h}\textrm{arctanh}\left[\left(\cosh\left(\frac{r_h\phi_{21}}{\ell}\right)+\sinh^2\left(\frac{r_h t_1}{\ell^2}\right)\right)\sech^2\left(\frac{r_h t_1}{\ell^2}\right)\tanh\left(\frac{r_h t_1}{\ell^2}\right)\right]\,.\label{extremization-EW-HM-adj}
	\end{align}
	Therefore, the length of the extremal curve $\Sigma_{AB}$ is obtained to be
	\begin{align}\label{Geodesic-length-SR-adj2(i)}
		\cL(\Sigma_{AB})=\ell \log\left[\frac{4\ell^2}{r_h\epsilon}\sinh \left(\frac{r_h \phi_{21}}{2\ell}\right) \sqrt{1+\sech^2\left(\frac{r_h t_1}{\ell^2}\right)\sinh^2 \left(\frac{r_h \phi_{21}}{2\ell}\right)}\right]
	\end{align}
	The above expression may also be computed by using the Kruskal-like coordinates $(s,y)$. A derivation of this is given in \cref{geodesic-length-s-y-adj2(i)}.
	Another approach to obtain the above expression is to directly use the prescription \eqref{EWCS-Adj} in the embedding coordinates. However, note that the \eqref{EWCS-Adj} applies only for cases in which the endpoints of the RT surface on which the EWCS ends are on the asymptotic boundary. To utilize the formula \eqref{EWCS-Adj}, one should choose the endpoint $X_3^A$ as 
	\begin{align*}
		X_3^A=\left(\phi_1,\frac{-t_1+\frac{i\beta}{2}}{\ell}\right),
	\end{align*}
	namely the endpoint of (extension of) the HM surface on the left asymptotic boundary. Note that upon utilizing the Brown-Henneaux relation and the Wick rotation $\tau_1= i t_1$, the reflected entropy computed from both perspectives matches identically.
	
	Furthermore, the radial location of the endpoint of the EWCS $\Sigma_{AB}$ on the EOW brane may be obtained by substituting \cref{extremization-EW-HM-adj} in \cref{HM-equation} as follows
	\begin{align}
		\hat{r}=\frac{r_h\sech^2\left(\frac{r_h t_1}{\ell^2}\right)\left[\sinh^2\left(\frac{r_h t_1}{\ell^2}\right)+\cosh \left(\frac{r_h \phi_{21}}{\ell}\right)\right]}{2\sinh\left(\frac{r_h\phi_{21}}{2\ell}\right)\sqrt{\cosh \left(\frac{2 r_h t_1}{\ell^2}\right)+\cosh \left(\frac{r_h \phi_{21}}{\ell}\right)}}
	\end{align}
	In particular, on the initial time slice of Lorentzian evolution, we have
	\begin{align}
		\hat{r}(t_1=0)=r_h\coth\left(\frac{r_h\phi_{21}}{\ell}\right)\label{EW-on-HM-r-adj}
	\end{align}
	which is greater than $r_h$ for any subsystem size. Therefore, this surface does not necessarily cross the horizon. In fact, it may be shown that the EWCS in this phase never crosses the horizon and its extension lands on the asymptotic boundary. This constitutes a new probe for behind-the-horizon geometry, which never crosses the horizon.

	\subsubsection*{Phase-II}

	\paragraph*{The boundary perspective:} In this reflected entropy phase, consider that both the subsystems are large, hence the EWCS lands on the EOW brane and divide the EE island into two parts.  
	In the boundary description, the effective reflected entropy between quantum matter fields in \cref{SR-bdy} may be obtained by utilizing the following twist field correlator 
	\begin{align}
		&S^{\text{eff}}_{R}(A\cup I_{S_{R}}(A):B \cup I_{S_{R}}(B))= \lim_{{m,n} \to 1}\frac{1}{1-n}\notag\\
		&\quad\quad  \times\log\frac{\langle \sigma_{g_A}(w_1,\bar{w}_1)\sigma_{g^{-1}_A}(w^{I}_1,\bar{w}^{I}_1)\sigma_{g^{}_A g^{-1}_B}(w^{}_2,\bar{w}^{}_2)\sigma_{g^{-1}_B}(w^{}_3,\bar{w}^{}_3)\sigma_{g^{}_B}(w^{I}_3,\bar{w}^{I}_3)\sigma_{g^{}_B g^{-1}_A}(w^{I}_b,\bar{w}^{I}_b)\rangle_{mn}}{\langle \sigma_{g_m}(w_1,\bar{w}_1)\sigma_{g^{-1}_m}(w^{I}_1,\bar{w}^{I}_1)\sigma_{g^{-1}_m}(w^{}_3,\bar{w}^{}_3)\sigma_{g^{}_m}(w^{I}_3,\bar{w}^{I}_3)\rangle^{n}_{m}}.\label{SR-eff-Adj2}
	\end{align}
	In the large central charge limit, the numerator of the above equation may be factorized into three two-point twist field correlators as  \cite{Chandrasekaran:2020qtn,Li:2020ceg}
	\begin{align}
		&\langle \sigma_{g_A}(w_1,\bar{w}_1)\sigma_{g^{-1}_A}(w^{I}_1,\bar{w}^{I}_1)\sigma_{g^{}_A g^{-1}_B}(w^{}_2,\bar{w}^{}_2)\sigma_{g^{-1}_B}(w^{}_3,\bar{w}^{}_3)\sigma_{g^{}_B}(w^{I}_3,\bar{w}^{I}_3)\sigma_{g^{}_B g^{-1}_A}(w^{I}_b,\bar{w}^{I}_b)\rangle_{mn}\notag\\
		& = \langle \sigma_{g_A}(w_1,\bar{w}_1)\sigma_{g^{-1}_A}(w^{I}_1,\bar{w}^{I}_1)\rangle_{mn}\langle \sigma_{g^{-1}_B}(w^{}_3,\bar{w}^{}_3)\sigma_{g^{}_B}(w^{I}_3,\bar{w}^{I}_3)\rangle_{mn}\langle \sigma_{g^{}_A g^{-1}_B}(w^{}_2,\bar{w}^{}_2)\sigma_{g^{}_B g^{-1}_A}(w^{I}_b,\bar{w}^{I}_b)\rangle_{mn}.
	\end{align}
	The first two twist field correlators of the above equation cancel with a similar factorization in the denominator and hence \cref{SR-eff-Adj2} may be rewritten as follows
	\begin{align}
		&S^{\text{eff}}_{R}(A\cup I_{S_{R}}(A):B \cup I_{S_{R}}(B))= \lim_{{m,n} \to 1}\frac{1}{1-n}\log \langle \sigma_{g^{}_A g^{-1}_B}(w^{}_2,\bar{w}^{}_2)\sigma_{g^{}_B g^{-1}_A}(w^{I}_b,\bar{w}^{I}_b)\rangle_{mn}.
	\end{align}
	Here $w^{I}_b=\left(\phi_2,\frac{\tau_1+\tau^{I}_b}{\ell}\right)$ is the intersection point between the EWCS and the EOW brane. Note that the field theory is defined on a hybrid manifold with the topology of conformal cylinders and hence the computation of the above twist field correlator is not straightforward. Therefore we need to map this twist field correlator to the complex plane twist field correlator which may be done by using \cref{CFT_DHP,UHP-map}. Utilizing these maps the above equation may be written as follows
	\begin{align}
		&S^{\text{eff}}_{R}(A\cup I_{S_{R}}(A):B \cup I_{S_{R}}(B))\notag\\ 
		&= \lim_{{m,n} \to 1}\frac{1}{1-n}\log \left(\tilde{\Omega} ~ \hat{\Omega}(z_2)\right)^{2 h_{AB}}  \left(\hat{\Omega}(z^{I}_b)\cdot\Omega(\tau^{I}_b)\right)^{2 h_{AB}} \langle \sigma_{g^{}_A g^{-1}_B}(z^{}_2,\bar{z}^{}_2)\sigma_{g^{}_B g^{-1}_A}(z^{I}_b,\bar{z}^{I}_b)\rangle_{mn},
	\end{align}
	where the conformal factors are given in \cref{Conformal-factors,Conformal-factor-bath-cft}. Now using the form of the two-point function, restoring to the original coordinates and adding the area term given in \cref{Area-term}, the reflected entropy for this phase may be obtained as
	\begin{align}\label{sr2(ii)}
		S_{R}^{\text{bdy}}(A:B)= \frac{c}{3} \log\left[\frac{2 r_0 \ell^3 \left(1+\sin \frac{r^{}_{h} (\tau^{I}_b-\tau_1)}{\ell^2}\right)}{\epsilon ~\epsilon_b  r^{2}_{h} \sin\frac{r^{}_{h} \tau^{I}_b}{\ell^2}}\right]+\frac{c}{3}\log \sqrt{\frac{1+ T \ell }{1-T \ell }}.	
	\end{align}
	Extremizing the above equation over $\tau^{I}_b$, we get the brane conformal time as
	\begin{align}
		\tau^{I}_b= \frac{ \pi \ell^2}{2 r_h}-\tau_1.
	\end{align}
	Substituting this in \cref{sr2(ii)}, the reflected entropy between two adjacent subsystems in this phase is given by
	\begin{align}
		S_{R}^{\text{bdy}}(A:B)=\frac{c}{3}\left(\log \left[\frac{2 \ell^2 }{r_h \epsilon }\cos \left(\frac{r_h \tau_1}{\ell^2}\right)\right]+\log \frac{2 r_0 \ell}{r_h \epsilon_b }+\log \sqrt{\frac{1+ T \ell }{1-T \ell }}\right).
	\end{align}

	\paragraph*{The bulk perspective:}
	\begin{figure}[ht]
		\centering
		\includegraphics[scale=1.2]{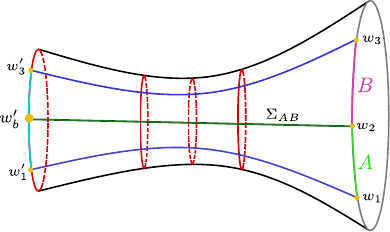}
		\caption{Diagrammatic illustration of the bulk EWCS between subsystems $A$ and $B$, shown as solid green curve.  } 
		\label{fig:adj2(ii)}
	\end{figure}
	
	In the bulk description, the curve $\Sigma_{AB}$ joins point $w_2$ to a point $w^{\prime}_b$ on the EOW brane, shown as the green curve in \cref{fig:adj2(ii)}.
	Recall that the effective reflected entropy in \cref{SR-bulk} reduces to $S^{\text{eff}}_R(I_{S_{R}}(A):I_{S_{R}}(B))$, which can be computed as follows
	\begin{equation}
		S^{\text{eff}}_{R}(\mathcal{A}:\mathcal{B})=S_{R}^\text{eff}(I_{S_{R}}(A):I_{S_{R}}(B))= \lim_{{m,n} \to 1}\frac{1}{1-n}\log \frac{   \langle \sigma_{g^{}_A}(w^{\prime}_1)\sigma_{g^{}_B g_A^{-1}}(w^{\prime}_b)\sigma_{g^{-1}_B}(w^{\prime}_3)\rangle_{\mathrm{BCFT}^{\bigotimes mn}}}{  \langle\sigma_{g^{}_m}(w^{\prime}_1)\sigma_{g^{-1}_m}(w^{\prime}_3)\rangle^n_{\mathrm{BCFT}^{\bigotimes m}}},
	\end{equation}
	where $w^{\prime}_i= (\phi_i, \tau^{\prime}_1)$.  
	As discussed earlier, the brane is conformally equivalent to a cylinder where computation of the above twist field correlator is not straightforward. 
	So it is necessary to transform this cylinder to upper half plane (UHP) which may be done by using the conformal transformation given in \cref{UHP-map}. 
	Utilizing this the above expression may be rewritten as follows
	\begin{align}
		S^{\text{eff}}_{R}(\mathcal{A}:\mathcal{B})= \lim_{{m,n} \to 1}\frac{1}{1-n}\log \frac{   \left(\hat{\Omega}(z^{\prime}_b) ~ \Omega(\tau^{\prime}_b)\right)^{2 h_{i}}   \langle \sigma_{g^{}_A}(z^{\prime}_1)\sigma_{g^{}_B g_A^{-1}}(z^{\prime}_b)\sigma_{g^{-1}_B}(z^{\prime}_3)\rangle_{\mathrm{BCFT}^{\bigotimes mn}}}{  \langle\sigma_{g^{}_m}(z^{\prime}_1)\sigma_{g^{-1}_m}(z^{\prime}_3)\rangle^n_{\mathrm{BCFT}^{\bigotimes m}}},
	\end{align}
	where the conformal factors given in \cref{Conformal-factors}.
	The above BCFT twist field correlators may be  expanded into two possible channel: the boundary operator expansion (BOE) and operator product expansion (OPE).  As discussed in \cref{subsec;review EE}, the bulk (OPE) channel expansion of the above twist correlator never dominates in the large central charge limit. 
	Utilizing the BOE channel, the twist field correlators may be factorized into one-point twist field correlator on the BCFT and after cancelling the twist field correlators corresponding to points $z^{\prime}_1$ and $z^{\prime}_3$, the resulting expression may be written  as follows
	\begin{align}\label{SR(eff.)_adj2(ii)}
		&S^{\text{eff}}_{R}(\mathcal{A}:\mathcal{B})=S_{R}^\text{eff}(I_{A}:I_{B})\notag\\
		&= \lim_{{m,n} \to 1}\frac{1}{1-n}\log  \left(\hat{\Omega}(z^{\prime}_b) ~ \Omega(\tau^{\prime}_b)\right)^{2 h_{AB}} \langle\sigma_{g^{}_B g_A^{-1}}(z^{\prime}_b)\rangle_{\mathrm{BCFT}^{\bigotimes mn}}.
	\end{align}
	Utilizing the form of one-point twist field correlator in the UHP and the appropriate conformal factors, we may obtain the effective reflected entropy as
	\begin{align}
		S^{\text{eff}}_{R}(\mathcal{A}:\mathcal{B})=\frac{c}{3}\log\left(\frac{2r_0\ell}{r_h\epsilon_b}\right)=\frac{c}{3}\log\left(\frac{2\ell}{\epsilon_b\sqrt{1-T^2}}\right).\,\label{SR-defect-contribution}
	\end{align}
	Furthermore, the area of the curve $\Sigma_{AB}$ may be obtained by following the same approach detailed in \cref{subsec;review EE}, leading to the result \eqref{Extremal-length}.
	Therefore, the reflected entropy for this phase in the bulk description may be obtained as 
	\begin{align}
		S^{\text{bulk}}_{R}(\mathcal{A}:\mathcal{B})=\frac{1}{2 G_N}\left(\log\left[ \frac{2\ell^2}{r_h\epsilon}\cosh \left(\frac{r_h t_1}{\ell^2}\right)\right]+\log \left(\frac{2 r_0 \ell}{r_h \epsilon_b }\right)+\log \sqrt{\frac{1+ T \ell }{1-T \ell }}\right).
	\end{align}

	\subsubsection*{Phase-III}
	\begin{figure}[H]
		\centering
		\includegraphics[scale=1.2]{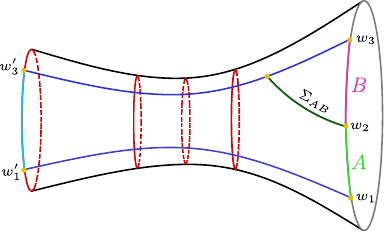}
		\caption{Diagrammatic illustration of the bulk EWCS between subsystems $A$ and $B$, depicted as solid green curve.  } 
		\label{fig:adj2(iii)}
	\end{figure}
	
	For this phase assume that the subsystem $B$ is smaller than the subsystem $A$, therefore the EWCS lands on the extremal surface corresponding to the points $w_3$ and $w^{\prime}_3$, depicted as solid green curve in \cref{fig:adj2(iii)}. The reflected entropy in the boundary description may be obtained by exchanging $\phi_{1}$ and $\phi_{3}$ in \cref{SR_adj_2(i)} as follows
	\begin{align}\label{SR_adj2(iii)}
		S^{\text{bdy}}_R(A:B)= \frac{c}{3} \log \left[\frac{2 \sqrt{2} \ell^2}{r_h \epsilon}\sec \left(\frac{r_h \tau_1}{\ell^2}\right) \sinh \left(\frac{r_h \phi_{32}}{2 \ell}\right) \sqrt{\cos \left(\frac{2 r_h \tau_1}{\ell^2}\right)+\cosh \left(\frac{r_h \phi_{32}}{\ell}\right)}\right].
	\end{align}
	In the bulk perspective, the reflected entropy may be computed in a similar manner to \hyperref[SR_adj_EE2(i)]{phase-I} or simply by exchanging the points $\phi_1$ and $\phi_3$ which exactly matches with \cref{SR_adj2(iii)} upon utilizing the Brown-Henneaux relation and the Wick rotation.

	\subsection{Disjoint Subsystems}\label{sec:disjoint}
	In this subsection, we analyze the reflected entropy corresponding to two disjoint subsystems $A=\left[(\phi_1,\frac{\tau_1}{\ell}),(\phi_2,\frac{\tau_1}{\ell})\right]$ and $B=\left[(\phi_3,\frac{\tau_1}{\ell}),(\phi_4,\frac{\tau_1}{\ell})\right]$ on a time slice $\tau=\tau_1$ of the asymptotic boundary of the braneworld cosmology discussed in \cref{sec:review}.
	To compute the reflected entropy or the bulk EWCS, it is necessary to first identify the entanglement entropy phases for the two subsystems under consideration. Depending on the size and location of the subsystems, we identify two distinct phases of EE. In the following, we provide a detailed computation of reflected entropy from both boundary and bulk perspectives for these EE phases.
	We demonstrate a precise agreement between the two approaches, establishing the consistency of the results.
	
	In the following, we will often utilize the comprehensive expression for the EWCS between two disjoint intervals $A=[X_1,X_2]$ and $B=[X_3,X_4]$ on the asymptotic boundary, written in the embedding coordinates \cite{Kusuki:2019evw}: 
	\begin{align}\label{EWCS-Dis}
		E_W= \frac{1}{4 G_N} \cosh^{-1} \left(\frac{1+\sqrt{u}}{\sqrt{v}}\right),
	\end{align}
	where
	\begin{align}
		u=\frac{\zeta_{12}\zeta_{34}}{\zeta_{13}\zeta_{24}},\; v=\frac{\zeta_{14}\zeta_{23}}{\zeta_{13}\zeta_{24}}.
	\end{align}

	\subsubsection{Entanglement Entropy phase 1}\label{EE-dis-phase-1}
	\begin{figure}[ht]
		\centering
		\includegraphics[scale=1.2]{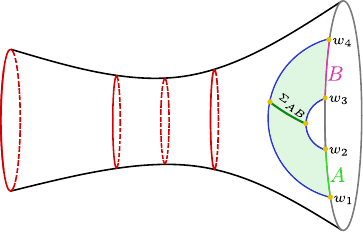}
		\caption{Schematic illustrating the EE phase $1$ when the RT surfaces for $A \cup B$ and the
			EWCS are represent as solid blue and green curve respectively. } 
		\label{fig:dis-1}
	\end{figure}
	In this EE phase, we consider that both the subsystems are small and  in close proximity to each other. As a result, the EE is given by the sum of the lengths of two dome-shaped RT surfaces shown by solid blue curves in \cref{fig:dis-1}. 
	From \cref{BTZ-invariant}, the EE for this configuration may be obtained as
	\begin{align}\label{dis-EE-1}	
		S^{(1)}_{AB}=\frac{1}{2 G_N}	\log \left[\left(\frac{2 \ell^2 }{r_h\epsilon }\right)^2\sinh \left(\frac{r_h \phi_{41}}{2 \ell}\right) \sinh \left(\frac{r_h \phi_{32}}{2\ell}\right)\right].
	\end{align}
	For this EE phase, we only have one phase of the reflected entropy or the bulk EWCS, shown as solid green curve in \cref{fig:dis-1}. From the boundary perspective, the reflected entropy may be computed using the four-point twist field correlator as
	\begin{align}
		S_R^{\text{bdy}}(A:B)= \lim_{{m,n} \to 1}\frac{1}{1-n}\log\frac{\langle \sigma_{g_A}(w_1, \bar{w}_1)\sigma_{g_A^{-1}}(w_2,\bar{w}_2)\sigma_{g_B}(w_3,\bar{w}_3)\sigma_{g^{-1}_B}(w_4,\bar{w}_4)\rangle_{mn}}{\langle \sigma_{g_m}(w_1,\bar{w}_1)\sigma_{g_m^{-1}}(w_2,\bar{w}_2)\sigma_{g_m}(w_3,\bar{w}_3)\sigma_{g^{-1}_m}(w_4,\bar{w}_4)\rangle^{n}_{m}}.
	\end{align}
	The bath BCFT$_2$ being defined on a cylinder, the computation of the above twist field correlator is not straightforward. It is necessary to transform the above four-point twist field correlator to the complex plane through the conformal map \eqref{CFT_DHP}:
	\begin{align}\label{SR_dis1(1)}
		S_R^{\text{bdy}}(A:B)= \lim_{{m,n} \to 1}\frac{1}{1-n}\log\frac{\langle \sigma_{g_A}(z_1,\bar{z}_1)\sigma_{g_A^{-1}}(z_2,\bar{z}_2)\sigma_{g_B}(z_3,\bar{z}_3)\sigma_{g^{-1}_B}(z_4,\bar{z}_4)\rangle_{mn}}{\langle \sigma_{g_m}(z_1,\bar{z}_1)\sigma_{g_m^{-1}}(z_2,\bar{z}_2)\sigma_{g_m}(z_3,\bar{z}_3)\sigma_{g^{-1}_m}(z_4,\bar{z}_4)\rangle^{n}_{m}}.
	\end{align}
	The four-point function on the complex plane may be expanded in terms of conformal blocks $\mathcal{F}\,,\bar{\mathcal{F}}$ as follows,
	\begin{align}
		\langle \sigma_{g_A}(z_1)\sigma_{g_A^{-1}}(z_2)\sigma_{g_B}(z_3)\sigma_{g^{-1}_B}(z_4)\rangle=\sum_{p}\mathcal{C}_{n,m}^2\,\mathcal{F}(mnc,h,h_p,\eta)\,\bar{\mathcal{F}}(mnc,h,h_p,\bar\eta)\,,
	\end{align}
	where $C_{n,m}$ is the OPE coefficient appearing in the three-point function, $\eta=\frac{(z_1-z_2)(z_3-z_4)}{(z_1-z_3)(z_2-z_4)}$ is the conformal cross-ratio. In the large central charge limit, the conformal block contributing to the four-point function on the complex plane is given by\footnote{As discussed in \cite{Dutta:2019gen}, the dominant contribution to the partial wave expansion is received from the block corresponding to the heavy operator $\sigma_{g^{}_Ag_B^{-1}}$.} \cite{Dutta:2019gen,Fitzpatrick:2014vua}
	\begin{align}
		\log\mathcal{F}(mnc,h,h_{AB},\eta)=-2h\log\eta+2h_{AB}\log\left[\frac{1+\sqrt{\eta}}{2\left(1-\sqrt{\eta}\right)}\right]\,.
	\end{align}
	Substituting the above expressions in \cref{SR_dis1(1)}, using the conformal dimensions \eqref{conformal dimensions} and subsequently taking the replica limit, the reflected entropy in the boundary description for this case may now be obtained as
	\begin{align}\label{SR-dis-1}
		S_R^{\text{bdy}}(A:B)= \frac{c}{6} \log \left[\frac{1+ \sqrt{\eta}}{1-\sqrt{\eta}}\right]+\frac{c}{6} \log \left[\frac{1+ \sqrt{\bar{\eta}}}{1-\sqrt{\bar{\eta}}}\right],
	\end{align}
	where the cross ratios $(\eta,\bar\eta)$ are given by
	\begin{align}
		\eta = \bar{\eta}= \sinh \left(\frac{r_h \phi_{21}}{2 \ell}\right) \csch\left(\frac{r_h \phi_{31}}{2 \ell}\right) \csch\left(\frac{r_h \phi_{42}}{2 \ell}\right) \sinh \left(\frac{r_h \phi_{43}}{2 \ell}\right).
	\end{align}
	In the bulk perspective, the EWCS is proportional to the length of the green geodesic $\Sigma_{AB}$ as depicted in \cref{fig:dis-1}. Now by utilizing the embedding coordinate given in \cref{embedding-r} for points $w_1$, $w_2$, $w_3$ and $w_4$ in \cref{EWCS-Dis}, the EWCS may be obtained as
	\begin{align}
		S_R^{\text{bulk}}(\mathcal{A}:\mathcal{B})= \frac{1}{2 G_N}\cosh ^{-1}\left[\frac{1+\sinh \left(\frac{r_h \phi_{21}}{2 \ell}\right) \csch\left(\frac{r_h \phi_{31}}{2 \ell}\right) \csch\left(\frac{r_h \phi_{42}}{2 \ell}\right) \sinh \left(\frac{r_h \phi_{43}}{2 \ell}\right)}{\sinh \left(\frac{r_h \phi_{32}}{2 \ell}\right) \csch\left(\frac{r_h \phi_{31}}{2 \ell}\right) \csch\left(\frac{r_h \phi_{42}}{2 \ell}\right) \sinh \left(\frac{r_h \phi_{41}}{2 \ell}\right)}\right].
	\end{align}
	Note that upon utilization of the Brown-Henneaux relation, we find that the reflected entropy computed in \cref{SR-dis-1} exactly matches with the above expression.

	\subsubsection{Entanglement entropy phase 2}\label{EE-dis-phase-2}
	\begin{figure}[ht]
		\centering
		\includegraphics[scale=1.2]{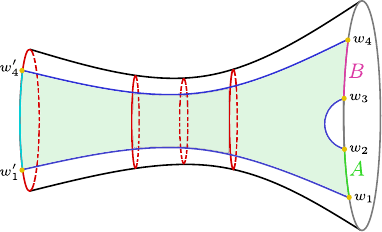}
		\caption{Schematic illustrating the EE phase $1$ when the RT surfaces for $A \cup B$ are represent as solid blue curves.  } 
		\label{fig:dis-2}
	\end{figure}
	
	In this EE phase, both the subsystems are large and away from each other, such that the EE corresponds to the sum of the length of a dome-shaped RT surface and two RT surface that crosses the horizon and end at the EOW brane, depicted as the blue curves in \cref{fig:dis-2}. Now utilizing \cref{DES-result}, the EE for this phase is given as follows
	\begin{align}
		S^{(2)}_{AB}=&\frac{1}{2 G_N}\log \left[\frac{2 \ell^2 }{r_h \epsilon }\sinh \left(\frac{r_h \phi_{32}}{2 \ell}\right)\right]\notag\\
		&+\frac{1}{2 G_N}\left(\log \left[\frac{2 \ell^2}{r_h \epsilon}\cosh \left(\frac{r_h t}{\ell^2}\right)\right]+\log \left(\frac{2 \ell}{\epsilon_b\sqrt{1- T^2 \ell^2}}\right)+\log \sqrt{\frac{1+ T \ell }{1-T \ell }}\right)
		\,,
	\end{align}
	where $\epsilon_b$ is the UV cut off on the EOW brane. In this EE phase, we identify three
	distinct phases for the reflected entropy or the bulk EWCS, depending on the subsystem size and their relative location. The computation of the reflected entropy for each phase, from both the boundary and bulk perspectives, is detailed in the following subsection.

	\subsubsection*{Phase-I}\label{SR-dis_2(i)}
	\paragraph*{The boundary perspective:} In this reflected entropy phase, we assume that the subsystem $A$ is smaller than the subsystem $B$, therefore the EWCS connects a dome-shaped RT surface to the extremal surface corresponding to the points $w_1$ and $w^{\prime}_1$. In the boundary description, the first term of \cref{SR-bdy} vanishes since there is no cross section on the EOW brane.
	So the reflected entropy in this phase reduces to $S^{\text{(eff)}}_{R}(A:B\cup I_{S_{R}}(B))$ which may be computed as follows
	\begin{align}\label{SR-dis2(i)-factor}
		&S^{\text{eff}}_{R}(A\cup I_{S_{R}}(A):B \cup I_{S_{R}}(B))= \lim_{{m,n} \to 1}\frac{1}{1-n}\notag\\
		&\quad\quad  \times\log\frac{\langle \sigma_{g_A}(w_1,\bar{w}_1)\sigma_{g_A^{-1}}(w_2,\bar{w}_2)\sigma_{g_B}(w_3,\bar{w}_3)\sigma_{g^{-1}_A}(w^{I}_1,\bar{w}^{I}_1)\sigma_{g^{-1}_B}(w^{}_4,\bar{w}^{}_4)\sigma_{g^{}_B}(w^{I}_4,\bar{w}^{I}_4)\rangle_{mn}}{\langle \sigma_{g_m}(w_1,\bar{w}_1)\sigma_{g_m^{-1}}(w_2,\bar{w}_2)\sigma_{g_m}(w_3,\bar{w}_3)\sigma_{g^{-1}_m}(w^{I}_1,\bar{w}^{I}_1)\sigma_{g^{-1}_m}(w^{}_4,\bar{w}^{}_4)\sigma_{g^{}_m}(w^{I}_4,\bar{w}^{I}_4)\rangle^{n}_{m}},
	\end{align}
	where $w^{I}_1=(\phi_{1},\frac{\tau_0+\tau^{I}_1}{\ell})$ is the intersection point between the EOW brane and the HM surface corresponding to the endpoint $w_1$. In the large central charge limit, the numerator of the above equation is factorized into a four-point and a two-point twist field correlator as 
	\begin{align}
		&\langle \sigma_{g_A}(w_1,\bar{w}_1)\sigma_{g_A^{-1}}(w_2,\bar{w}_2)\sigma_{g_B}(w_3,\bar{w}_3)\sigma_{g^{-1}_A}(w^{I}_1,\bar{w}^{I}_1)\sigma_{g^{-1}_B}(w^{}_4,\bar{w}^{}_4)\sigma_{g^{}_B}(w^{I}_4,\bar{w}^{I}_4)\rangle_{mn}\notag\\
		&\quad \quad =\langle \sigma_{g_A}(w_1,\bar{w}_1)\sigma_{g_A^{-1}}(w_2,\bar{w}_2)\sigma_{g_B}(w_3,\bar{w}_3)\sigma_{g^{-1}_A}(w^{I}_1,\bar{w}^{I}_1)\rangle_{mn} \times \langle\sigma_{g^{-1}_B}(w^{}_4,\bar{w}^{}_4)\sigma_{g^{}_B}(w^{I}_4,\bar{w}^{I}_4)\rangle_{mn},
	\end{align}
	The denominator of \cref{SR-dis2(i)-factor} admits a similar factorization. Hence the reflected entropy in this phase may be obtained by using the following twist field correlator
	\begin{align}
		S^{\text{(bdy)}}_{R}(A:B)=\lim_{{m,n} \to 1}\frac{1}{1-n}\log\frac{\langle \sigma_{g_A}(w_1,\bar{w}_1)\sigma_{g_A^{-1}}(w_2,\bar{w}_2)\sigma_{g_B}(w_3,\bar{w}_3)\sigma_{g^{-1}_A}(w^{I}_1,\bar{w}^{I}_1)\rangle_{mn}}{\langle \sigma_{g_m}(w_1,\bar{w}_1)\sigma_{g_m^{-1}}(w_2,\bar{w}_2)\sigma_{g_m}(w_3,\bar{w}_3)\sigma_{g^{-1}_m}(w^{I}_1,\bar{w}^{I}_1)\rangle^{n}_{m}}.
	\end{align} 
	Now by using \cref{CFT_DHP,UHP-map}, we may map the above twist field correlators to the complex plane.
	Utilizing the form of the four-point twist field correlator in the large central charge limit \cite{Dutta:2019gen,Fitzpatrick:2014vua} and substituting the value of the brane time given in \cref{brane-time-EE} (determined by the extremization of the EE for $A\cup B$) we may obtain the expression for the reflected entropy  for this phase in the boundary description identical to \cref{SR_dis1(1)} with the cross ratios $\eta$, $\bar{\eta}$ given as
	\begin{align}\label{SR_dis2(i)}
		&\eta= \csch\left(\frac{r_h \phi_{21}}{2 \ell}\right) \sinh \left(\frac{r_h \phi_{31}}{2 \ell}\right) \cosh \left(\frac{r_h (2 t_1+\ell  \phi_{21})}{2 \ell^2}\right)\sech \left(\frac{r_h (2 t_1+\ell  \phi_{31})}{2 \ell^2}\right),\notag\\
		&	\bar{\eta}= \csch\left(\frac{r_h \phi_{21}}{2 \ell}\right) \sinh \left(\frac{r_h \phi_{31}}{2 \ell}\right) \cosh \left(\frac{r_h (-2 t_1+\ell  \phi_{21})}{2 \ell^2}\right)\sech \left(\frac{r_h (-2 t_1+\ell  \phi_{31})}{2 \ell^2}\right).
	\end{align}
	Note that in the above expression we have used the Wick rotation $\tau_1= i t_1$, in order to render the cross ratios real.

	\paragraph*{The bulk perspective:}
	\begin{figure}[ht]
		\centering
		\includegraphics[scale=1.2]{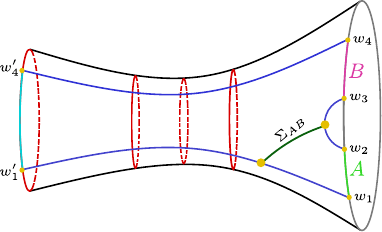}
		\caption{Schematic illustration of the bulk EWCS between subsystems $A$ and $B$, shown as solid green curve.  } 
		\label{fig:dis-2(i)}
	\end{figure}
	
	In the bulk description, the curve $\Sigma_{AB}$ joins an arbitrary point on the dome-shaped surface to another arbitrary point on the extremal surface (HM surface) corresponding to the endpoint $w_1$, shown as solid green curve in \cref{fig:dis-2(i)}. For this case 
	the second term of \cref{SR-bulk} vanishes owing to the fact that there is no contribution due to the brane matter as the entire island belongs to subsystem $B$. Hence, the reflected entropy between the bulk region $\mathcal{A}$ and $\mathcal{B}$ is determined only by the length of the minimal curve $\Sigma_{AB}$. 
	
	However the computation of the length of this geodesic by directly extremizing over its end points poses significant mathematical challenges because of the complexity of the calculations involved. Therefore we utilize the prescription given in \cref{EWCS-Dis} to obtain the reflected entropy for this phase in the bulk description.
	As discussed earlier in \cref{SR_adj_EE2(i)}, \cref{EWCS-Dis} 
	is applicable only when the endpoints of the RT surface, on which the EWCS ends are on the asymptotic boundary. 
	Consequently, to utilize \eqref{EWCS-Dis}, the endpoint $X_1^A$
	must be chosen as
	\begin{align*}
		X_1^A=\left(\phi_1,\frac{-t_1+\frac{i\beta}{2}}{\ell}\right),
	\end{align*}
	which corresponds to the endpoint of (extension of) the HM surface on the left asymptotic boundary.
	Now by using the embedding coordinates given in \cref{embedding-r}, the reflected entropy for this phase in the bulk description may be obtained as follows
	\begin{align}
		S^{\text{bulk}}_R(\mathcal{A}:\mathcal{B})= \frac{1}{2 G_N}\cosh ^{-1}\Bigg[&\frac{\sinh\left(\frac{r_h\phi_{21}}{2\ell}\right)}{\sinh\left(\frac{r_h\phi_{32}}{2\ell}\right)}\sqrt{1+\sech^2\left(\frac{r_h t_1}{\ell^2}\right)\sinh^2\left(\frac{r_h\phi_{31}}{2\ell}\right)}\notag\\&+\frac{\sinh\left(\frac{r_h\phi_{31}}{2\ell}\right)}{\sinh\left(\frac{r_h\phi_{32}}{2\ell}\right)}\sqrt{1+\sech^2\left(\frac{r_h t_1}{\ell^2}\right)\sinh^2\left(\frac{r_h\phi_{21}}{2\ell}\right)}\Bigg]
	\end{align}
	Upon using the Brown-Henneaux relation, the reflected entropy from both the perspective may be seen to match exactly. Furthermore, it is possible to deduce that the extremal surface $\Sigma_{AB}$ never crosses the event horizon and hence constitutes yet another probe of behind the horizon physics, without ever being able to cross the horizon. We defer the details of this calculation till \cref{appB}.

	\subsubsection*{Phase-II}\label{SR-dis_2(ii)}

\begin{figure}[ht]
		\centering
		\includegraphics[scale=1.2]{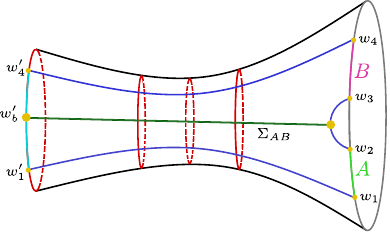}
		\caption{Diagrammatic illustration of the bulk EWCS between subsystems $A$ and $B$, depicted as solid green curve. } 
		\label{fig:dis-2(ii)}
	\end{figure}

	\paragraph*{The boundary perspective:} For this phase, the EWCS lands on the EOW brane and divide the EE island into two parts as depicted in \cref{fig:dis-2(ii)}. Now in the boundary description, the first term of \cref{SR-bdy} may be computed as follows
	\begin{align}\label{SR_dis2(i)_def}
		&S^{\text{eff}}_{R}(A\cup I_{S_{R}}(A):B \cup I_{S_{R}}(B))= \lim_{{m,n} \to 1}\frac{1}{1-n}\notag\\
		& \times\log\frac{\langle \sigma_{g_A}(w_1,\bar{w}_1)\sigma_{g^{-1}_A}(w^{I}_1,\bar{w}^{I}_1)\sigma_{g_A^{-1}}(w_2,\bar{w}_2)\sigma_{g_B}(w_3,\bar{w}_3)\sigma_{g^{-1}_B}(w^{}_4,\bar{w}^{}_4)\sigma_{g^{}_B}(w^{I}_4,\bar{w}^{I}_4)\sigma_{g^{}_B g^{-1}_A}(w^{I}_b,\bar{w}^{I}_b)\rangle_{mn}}{\langle \sigma_{g_m}(w_1,\bar{w}_1)\sigma_{g^{-1}_m}(w^{I}_1,\bar{w}^{I}_1)\sigma_{g_m^{-1}}(w_2,\bar{w}_2)\sigma_{g_m}(w_3,\bar{w}_3)\sigma_{g^{-1}_m}(w^{}_4,\bar{w}^{}_4)\sigma_{g^{}_m}(w^{I}_4,\bar{w}^{I}_4)\rangle^{n}_{m}},
	\end{align}
	where $w^{I}_b$ is the location of the island cross-section on the EOW brane.
	In the large central charge limit, the correlator on the numerator of the above expression may be factorized into two one-point twist field correlators and one three-point twist field correlator as
	\begin{align}\label{SR_dis2(ii)_Num}
		&\langle \sigma_{g_A}(w_1,\bar{w}_1)\sigma_{g^{-1}_A}(w^{I}_1,\bar{w}^{I}_1)\sigma_{g_A^{-1}}(w_2,\bar{w}_2)\sigma_{g_B}(w_3,\bar{w}_3)\sigma_{g^{-1}_B}(w^{}_4,\bar{w}^{}_4)\sigma_{g^{}_B}(w^{I}_4,\bar{w}^{I}_4)\sigma_{g^{}_B g^{-1}_A}(w^{I}_b,\bar{w}^{I}_b)\rangle_{mn}\notag\\
		&=\langle \sigma_{g_A}(w_1,\bar{w}_1)\sigma_{g^{-1}_A}(w^{I}_1,\bar{w}^{I}_1)\rangle_{mn} \times \langle\sigma_{g^{-1}_B}(w^{}_4,\bar{w}^{}_4)\sigma_{g^{}_B}(w^{I}_4,\bar{w}^{I}_4)\rangle_{mn}\notag\\
		&\qquad\qquad\qquad\qquad\qquad\qquad\qquad \times \langle\sigma_{g_A^{-1}}(w_2,\bar{w}_2)\sigma_{g_B}(w_3,\bar{w}_3)\sigma_{g^{}_B g^{-1}_A}(w^{I}_b,\bar{w}^{I}_b)\rangle_{mn},
	\end{align}
	with a similar factorization of the twist correlator in the denominator. 
	Now substituting \cref{SR_dis2(ii)_Num} into \cref{SR_dis2(i)_def}, the effective reflected entropy in \cref{SR-bdy} may be obtained as follows
	\begin{align}
		S^{\text{eff}}_{R}(A\cup I_{S_{R}}(A):B \cup I_{S_{R}}(B))=\lim_{{m,n} \to 1}\frac{1}{1-n}\log\ \frac{\langle\sigma_{g_A^{-1}}(w_2,\bar{w}_2)\sigma_{g_B}(w_3,\bar{w}_3)\sigma_{g^{}_B g^{-1}_A}(w^{I}_b,\bar{w}^{I}_b)\rangle_{mn}}{\langle \sigma_{g_m^{-1}}(w_2,\bar{w}_2)\sigma_{g_m}(w_3,\bar{w}_3)\rangle^{n}_{m}},
	\end{align}
	Using the conformal transformation given in \cref{CFT_DHP,UHP-map} to map the above twist field correlator to the complex plane twist field correlator, the above expression may be rewritten as 
	\begin{align}
		&S^{\text{eff}}_{R}(A\cup I_{S_{R}}(A):B \cup I_{S_{R}}(B))\notag\\
		&\quad \quad=\lim_{{m,n} \to 1}\frac{1}{1-n}\log  \frac{\left(\hat{\Omega}(z^{I}_b)\cdot\Omega(\tau^{I}_b)\right)^{2 h_{AB}} \langle\sigma_{g_A^{-1}}(z_2,\bar{z}_2)\sigma_{g_B}(z_3,\bar{z}_3)\sigma_{g^{}_B g^{-1}_A}(z^{I}_b,\bar{z}^{I}_b)\rangle_{mn}}{\langle \sigma_{g_m^{-1}}(z_2,\bar{z}_2)\sigma_{g_m}(z_3,\bar{z}_3)\rangle^{n}_{m}},
	\end{align}
	where the conformal factor is given in \cref{Conformal-factors} with $\tau^{\prime}_b= \tau_0+\tau^{I}_b$ is the brane conformal time and conformal factors corresponding to the points $z_1$ and $z_2$ cancel in numerator and denominator. Now by utilizing the form of the three and two point function and appropriate conformal factors, the above expression may be written as 
	\begin{align}
		S^{\text{eff}}_{R}(A\cup I_{S_{R}}(A):B \cup I_{S_{R}}(B))=&\frac{c}{3} \log \left[\frac{2 \ell r^{}_0   }{r^{}_h \epsilon_b }\csch\left(\frac{r^{}_h \phi_{32}}{2\ell}\right)\right]+\frac{c}{6} \log\csc\left(\frac{r_h \tau^{I}_b}{\ell^2}\right) \notag\\
		&+\frac{c}{12} \log \left[\sin\frac{r_h(\tau^{I}_b-\tau_1)}{\ell^2}-\cosh \frac{r_h (\phi_2-\phi_b)}{\ell}\right] \notag\\
		&+\frac{c}{12}\log \left[\sin \frac{r_h(\tau^{I}_b-\tau_1)}{\ell^2}-\cosh \frac{r_h(\phi_3-\phi_b)}{\ell}\right].
	\end{align}
	Here $(\phi_b,t_b)$ are the coordinates of the island cross section on the EOW brane. 
	As the area of the island cross-section is given by a constant \eqref{Area-term}, in order to obtain the reflected entropy we may extremize the above expression over $\phi_b$ and $\tau^{I}_b$, leading to the solutions
	\begin{align}
		\phi_b= \frac{\phi_{2}+\phi_{3}}{2} ~~,~~ \tau^{I}_b= \frac{2 \ell^2}{r_{h}} \tan^{-1} \sqrt{\displaystyle\frac{\cosh \left(\frac{r_h\phi_{32}}{2 \ell}\right)-\sin\left(\frac{r_{h}\tau_1}{\ell^2}\right)}{\cosh \left(\frac{r_h\phi_{32}}{2 \ell}\right)+\sin\left(\frac{r_{h}\tau_1}{\ell^2}\right)}}~~.
	\end{align}
	Substituting the extremal values and subsequently using the Wick rotation $\tau_1= i t_1$ in the resulting expression, the reflected entropy for this phase in the boundary description may be obtained as
	\begin{align}
		S^{\text{bdy}}_R(A:B)&=\frac{c}{3} \log \left[\frac{2 \ell r^{}_0  }{r^{}_h \epsilon_b }\csch\left(\frac{r^{}_h \phi_{32}}{2\ell}\right)\right]+\frac{c}{3} \log \sqrt{\frac{1+ T \ell }{1-T \ell }}\notag\\
		&+\frac{c}{6} \log \Bigg[1+2 \cosh \left(\frac{r_h t_1 }{\ell^2}\right) \sqrt{\cosh^2 \left(\frac{r_h t_1}{\ell^2}\right)+\sinh^2\left(\frac{r_h \phi_{32}}{2\ell}\right)}\notag\\
		&\qquad\qquad\qquad\qquad\qquad\qquad+2 \cosh \left(\frac{2 r_h t_1}{\ell^2}\right)+\cosh \left(\frac{r_h \phi_{32}}{\ell}\right)\Bigg].
	\end{align}

	\paragraph*{The bulk perspective:} 
	
	In the bulk description, the curve $\Sigma_{AB}$ connects the dome-shaped RT surface to an arbitrary point $w^{\prime}_b$, shown by the green curve in \cref{fig:dis-2(ii)}. 
	The effective reflected entropy between the two regions $I_{S_R}(A)$ and $I_{S_R}(B)$ is explicitly given in \cref{SR-defect-contribution}.
	The second term of \cref{SR-bulk} is proportional to the geodesic length of the curve $\Sigma_{AB}$ which may be computed by utilizing the embedding coordinates corresponding to the points $w_2$, $w_3$ and $w^{\prime}_b$ in \cref{EWCS-Adj} as 
	\begin{align}\label{length_dis2(ii)}
		L= \cosh ^{-1} \Bigg[\frac{\csch \frac{r_h \phi_{32}}{2 \ell}}{r_h} &\sqrt{ \sqrt{1-\frac{r_h^2}{r_b^2}} ~\cosh \frac{ r_h (t_1-t^{\prime}_b)}{\ell^2}- \cosh \frac{r_h (\phi_b-\phi_2)}{\ell}}\notag\\
		&\times \sqrt{\sqrt{1-\frac{r_h^2}{r_b^2}} ~\cosh \frac{r_h (t_1-t^{\prime}_b)}{\ell^2}- \cosh \frac{r_h (\phi_3-\phi_b)}{\ell}}\Bigg].
	\end{align}
	Now using \cref{brane-trajectory} we can write $r_b$ in terms of $t^{\prime}_b$ and then the EWCS may be obtained by extremizing the resulting expression over $\phi_b$ and $t^{\prime}_b$. The extremal value of $\phi_b$ and $t^{\prime}_b$ are then given as
	\begin{align}
		\phi_b= \frac{\phi_2+\phi_3}{2}, \quad\quad t^{\prime}_b= \frac{ \ell^2}{r_h}\arctanh \frac{\sinh\frac{r_h t_1}{\ell^2}}{T \ell} \sqrt{\frac{2}{\cosh \frac{2 r_h t_1}{\ell^2}+ \cosh\frac{r_h \phi_{32}}{\ell}}}.
	\end{align}
	Substituting these extremized value in \cref{length_dis2(ii)}, the second term of \cref{SR-bulk} may be written as 
	\begin{align}\label{area_dis2(ii)}
		\text{Area}[\Sigma_{AB}]&= \cosh^{-1} \left[\frac{\csch\frac{r_h \phi_{32}}{2 \ell}}{\sqrt{1- T^2 \ell^2}}\left(T \ell \cosh \frac{r_h t_1}{\ell^2}+\sqrt{\cosh^2\frac{r_h t_1}{\ell^2}+ \sinh^2\frac{r_h \phi_{32}}{2 \ell}}\right)\right]\notag\\&= \cosh^{-1} \left[\sqrt{1+\cosh^2\frac{r_h t_1}{\ell^2}\csch^2\frac{r_h \phi_{32}}{2 \ell}}\right]+ \cosh^{-1}\frac{1}{\sqrt{1- T^2 \ell ^2}}.
	\end{align}
	where, in the second equality, we have utilized the identity 
	\begin{align*}
		\cosh^{-1}(x)+\cosh^{-1}(y)=\cosh^{-1}\left(xy+\sqrt{(x^2-1)(y^2-1)}\right)\,.
	\end{align*}
	The geodesic length of the curve $\Sigma_{AB}$ may also be computed by using the Kruskal-like coordinates $(s,y)$ (cf. \cref{AppA}).
	Now by adding \cref{SR-defect-contribution,area_dis2(ii)} and using the Wick rotation $\tau_1= i t_1$, the reflected entropy for this phase in the bulk description may be obtained as
	\begin{align}
		S^{\text{bulk}}_{R}(\mathcal{A}:\mathcal{B})&=\frac{1}{2 G_N}\log\left(\frac{2r_0\ell}{r_H\epsilon_b}\right)+\frac{1}{2G_N} \log \sqrt{\frac{1+ T \ell}{1- T \ell}}\notag\\
		&+ \frac{1}{2G_N}\log \left[\csch\frac{r_h \phi_{32}}{2 \ell}\left( \cosh \frac{r_h t_1}{\ell^2}+\sqrt{\cosh^2\frac{r_h t_1}{\ell^2}+ \sinh^2\frac{r_h \phi_{32}}{2 \ell}}\right)\right].
	\end{align}
	Note that upon utilizing the Brown-Henneaux relation, the expression of the reflected entropy matches exactly from both the perspectives.

	\subsubsection*{Phase-III}
	\begin{figure}[ht]
		\centering
		\includegraphics[scale=1.2]{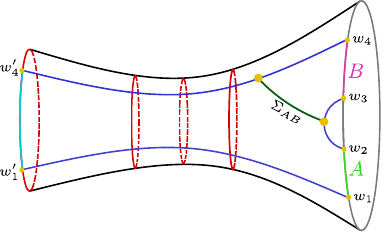}
		\caption{Diagrammatic illustration of the bulk EWCS between subsystems $A$ and $B$, depicted as solid green curve.   } 
		\label{fig:dis-2(iii)}
	\end{figure}
	For this reflected entropy phase, we consider that the subsystem $B$ is smaller than the subsystem $A$, hence the EWCS lands on the extremal surface corresponding to the points $w_4$ and $w^{\prime}_4$, shown as green geodesic in \cref{fig:dis-2(iii)}. Now the expression of the reflected entropy in the boundary description is identical to \cref{SR_dis1(1)} where the cross ratios $\eta$, $\bar{\eta}$ may be obtained 
	by interchanging $\phi_{1}$ and $\phi_{4}$ in \cref{SR_dis2(i)} as
	\begin{align}\label{SR_dis2(iii)}
		&\eta= \csch\left(\frac{r_h \phi_{42}}{2 \ell}\right) \sinh \left(\frac{r_h \phi_{43}}{2 \ell}\right) \cosh \left(\frac{r_h (2 t_1+\ell  \phi_{42})}{2 \ell^2}\right)\sech \left(\frac{r_h (2 t_1+\ell  \phi_{43})}{2 \ell^2}\right),\notag\\
		&	\bar{\eta}= \csch\left(\frac{r_h \phi_{42}}{2 \ell}\right) \sinh \left(\frac{r_h \phi_{43}}{2 \ell}\right) \cosh \left(\frac{r_h (-2 t_1+\ell  \phi_{42})}{2 \ell^2}\right)\sech \left(\frac{r_h (-2 t_1+\ell  \phi_{43})}{2 \ell^2}\right).
	\end{align}
	The bulk computation in this case may be performed in a manner similar to the \hyperref[SR-dis_2(i)]{phase-I} and the reflected entropy is obtained by interchanging $\phi_{1}$ and $\phi_{4}$ which precisely matches with the reflected entropy computed from the boundary description, when the Brown-Henneaux relation is used.

	\section{Time evolution of the reflected entropy}\label{sec:Plot}
	In this section, we explore the time evolution of reflected entropy for both adjacent and disjoint subsystems, as introduced earlier.  Additionally, we discuss the difference between the reflected entropy and the mutual information, which was termed as the Markov gap in \cite{Hayden:2021gno}. It has been illustrated that the Markov gap is bounded by the fidelity of a Markov recovery process associated with the purification of the mixed state under consideration.
	The authors in \cite{Hayden:2021gno} provide a geometric interpretation of the Markov gap in terms of the number of non-trivial boundaries of the EWCS. In the context of AdS$_3$/CFT$_2$, it was established that 
	\begin{align}\label{Markov gap}
		S_{R}(A: B)-I(A: B) \geq \frac{\log (2) \ell_{\text {AdS }}}{2 G_{N}} \times(\# \text { of boundaries of EWCS})+\mathcal{O}\left(\frac{1}{G_{N}}\right).
	\end{align}
	In the following, we find that for all cases where the bulk EWCS has no non-trivial boundaries, the Markov gap vanishes identically, indicating the possibility of a perfect Markov recovery process.

	\subsection{Adjacent subsystems}
	In this subsection, we explain the various phase transitions in reflected entropy over time for two adjacent subsystems and also analyse the Markov gap in this context. 
	In order to investigate the Markov gap, we first need to determine the mutual information phases between two adjacent subsystems under consideration.
	In the present scenario, we identify five distinct mutual information phases, depending on the subsystems size, which are given as follows
	\begin{align}
	I(A:B)=\begin{cases}
		\displaystyle&\frac{c}{3}\log\left[\frac{2\ell^2}{r_h \epsilon}\,\frac{\sinh \left(\frac{r_h \phi_{21}}{2 \ell}\right) \sinh \left(\frac{r_h \phi_{32}}{2 \ell}\right)}{\sinh\left(\frac{r_h \phi_{31}}{2 \ell}\right)}\right],\notag\\
		&\frac{c}{3}\left(\log\left[\frac{2 \ell^2}{r_h\epsilon }\sinh \left(\frac{r_h \phi_{21}}{2 \ell}\right) \sinh \left(\frac{r_h \phi_{32}}{2\ell}\right)\right]-\log \frac{2 r_0 \ell \cosh \frac{r_h t_1}{\ell^2}}{r_h \epsilon_b } -\log \sqrt{\frac{1+ T \ell }{1-T \ell }}\right),\notag\\
		&\frac{c}{3}\log\left(\frac{2 \ell^2 }{r_h\epsilon }\sinh \left(\frac{r_h \phi_{21}}{2 \ell}\right)\right),\\
		&\frac{c}{3}\log\left(\frac{2 \ell^2 }{r_h\epsilon }\sinh \left(\frac{r_h \phi_{32}}{2 \ell}\right)\right),\notag\\
		&\frac{c}{3}\left(\log \frac{2 \ell^2 \cosh \frac{r_h t_1}{\ell^2}}{r_h \epsilon }+\log \frac{2 r_0 \ell}{r_h \epsilon_b }+\log \sqrt{\frac{1+ T \ell }{1-T \ell }}\right).
	\end{cases}
	\end{align}
	
	\subsection*{Case-I}
	\begin{figure}[ht]
		\centering
		\begin{subfigure}[b]{0.45\linewidth}
			\centering
			\includegraphics[scale=.35]{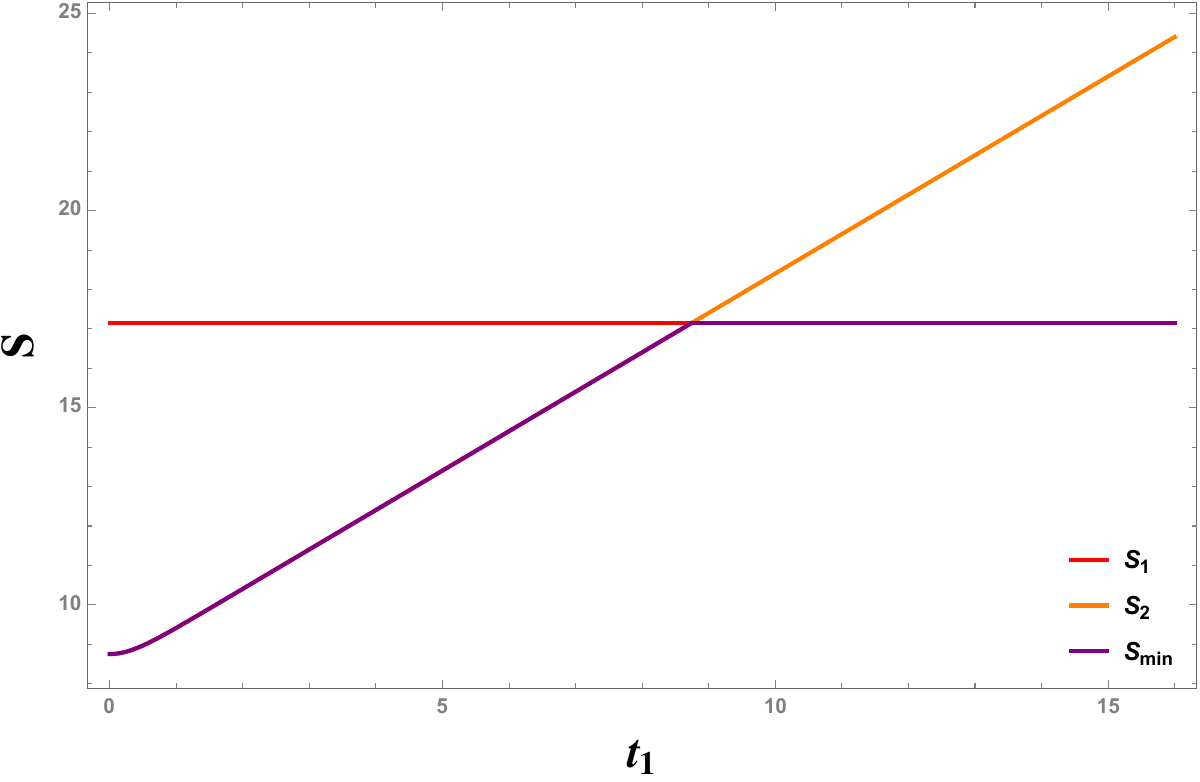}
			\caption{ }
			\label{EE-adjacent2}
		\end{subfigure}
		\hfill
		\begin{subfigure}[b]{0.45\linewidth}
			\centering
			\includegraphics[scale=.34]{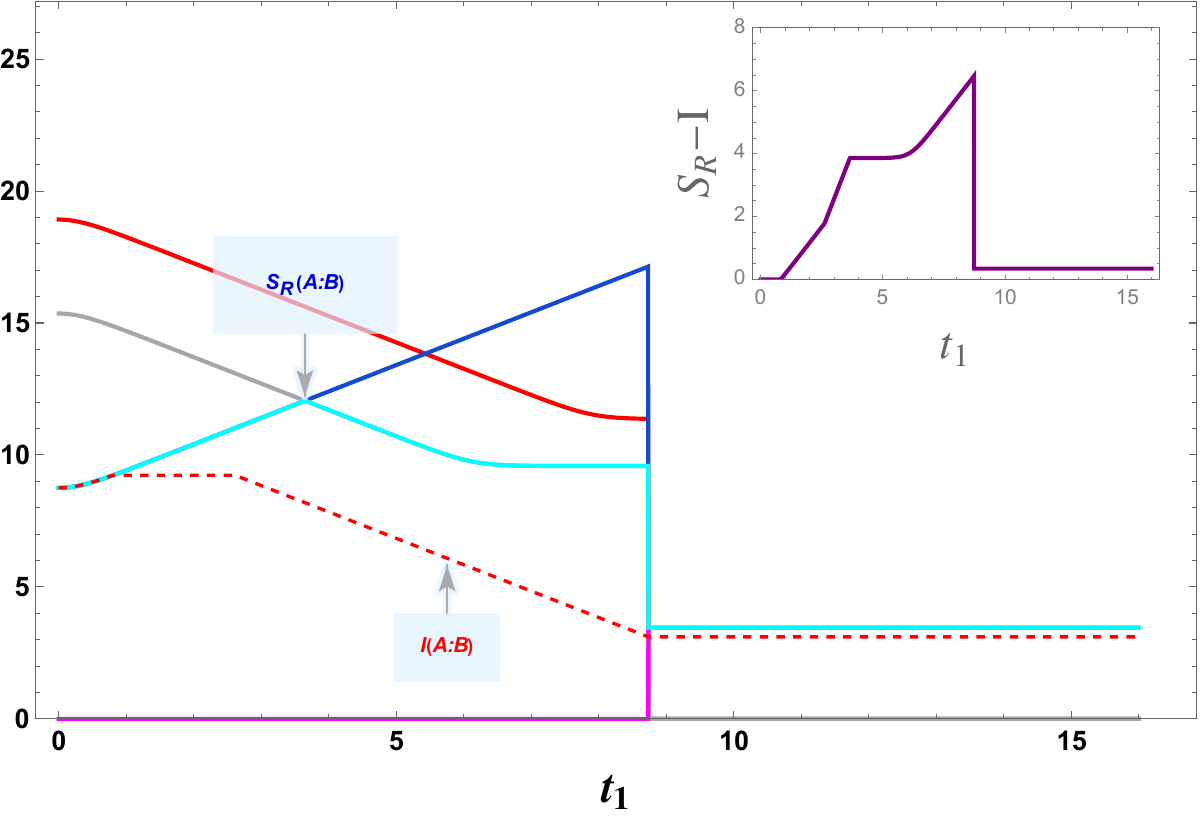}
			\caption{}
			\label{adjacent SR2}
		\end{subfigure}
		\caption{(a) EE for two adjacent subsystems $A \cup B$ vs time graph. Here purple curve indicates the minimum	EE among two phases. (b) Reflected entropy between subsystem $A$ and $B$ as a function of time. Here cyan curve shows minimum $S_R$ and red dashed curve is mutual information. (Both graphs are in units of $c$). The inset graph represents the difference between $S_R$ and mutual information. These plots are obtained with $\ell=1, T=.95, r_H=2 \ell, \epsilon=.001, \epsilon_b=.001, \phi_1=\frac{\pi}{6}, \phi_2=5.2 \pi, \phi_3=9.1 \pi.$ }
	\end{figure}
	
	The entanglement entropy (EE) phase transition between \hyperref[EE-adj2]{phase-II} and \hyperref[EE-adj1]{phase-I} may be understood by analysing the separation between the points $w_1$ and $w_3$. When these points are significantly far apart, the system undergoes a transition from one phase to another, as depicted in \cref{EE-adjacent2}. The transition time between these phases is given as
	\begin{align}\label{adj-EE-trans.-time}
		T^{\text{adj}}_{E}=\frac{\ell^2}{r_h}	\cosh ^{-1}\left(\frac{\epsilon_b(1- T \ell ) \sinh \frac{r_h \phi_{31}}{2 \ell}}{2 \ell^2}\right).
	\end{align}
	We now analyse the time evolution of the reflected entropy across different entanglement entropy (EE) phases, focusing on the scenario where subsystem $B$ is smaller than subsystem $A$.
	These reflected entropy phase transition is shown in \cref{adjacent SR2}.  
	Initially in the \hyperref[EE-adj2]{phase-II},
	the reflected entropy increases with time as the EWCS is the HM surface, which crosses the horizon and end at the EOW brane. 
	However, after a certain time $T^{\text{adj}}_{S_{R}}$, the reflected entropy begins to decrease and eventually stabilizes, remaining constant until the transition time $T^{\text{adj}}_{E}$ 
	as the EWCS lands on the HM surface corresponding to the point $w_3$ and $w^{\prime}_3$. In our computation, we have also shown that for this reflected entropy phase, the EWCS never crosses the horizon and decreases with time. The transition time between these reflected entropy phases is given as
	\begin{align}
		T^{\text{adj}}_{S_{R}}=\frac{\ell^2}{r_h}   \cosh ^{-1}\left(\frac{\epsilon_b(1-T \ell )}{\sqrt{2}\ell}\sinh \frac{r_h \phi_{21}}{2 \ell}\sqrt{ \left[1+\sqrt{1+\left[\frac{2 \ell}{\epsilon_b(1-T \ell )}\right]^2}\right]} \right).
	\end{align}
	Finally, in \hyperref[EE-adj1]{phase-I}, the reflected entropy saturates to a constant value.
	Now to investigate the Markov gap, we also plot the mutual information, shown as dashed red lines in \cref{adjacent SR2}.
	From the inset plot, we observe that the Markov gap initially vanishes, as both the bulk EWCS and mutual information are determined by the Hartman-Maldacena (HM) surface, which has no non-trivial boundaries.
	However, after some time, within the same reflected entropy phase, the Markov gap becomes non zero as the mutual information undergoes a phase transition. This observation appears to contradict the geometric interpretation of the Markov gap given in \cref{Markov gap}, suggesting a critical reassessment of this issue in the context of the KR braneworld scenario.
	Subsequently with time this gap increases to a value greater than $\frac{c}{3}\log2$ until the transition time $T^{\text{adj}}_{E}$ as the bulk EWCS has one non-trivial boundary and the mutual information decreases over time due to the phase transition in the mutual information, in the corresponding phase. 
	Finally, after the transition time $T^{\text{adj}}_{E}$, the Markov gap saturates to the lower bound mentioned earlier in \cref{Markov gap}.

	\subsection*{Case-II}

	\begin{figure}[ht]
		\centering
		\begin{subfigure}[b]{0.45\linewidth}
			\centering
			\includegraphics[scale=.35]{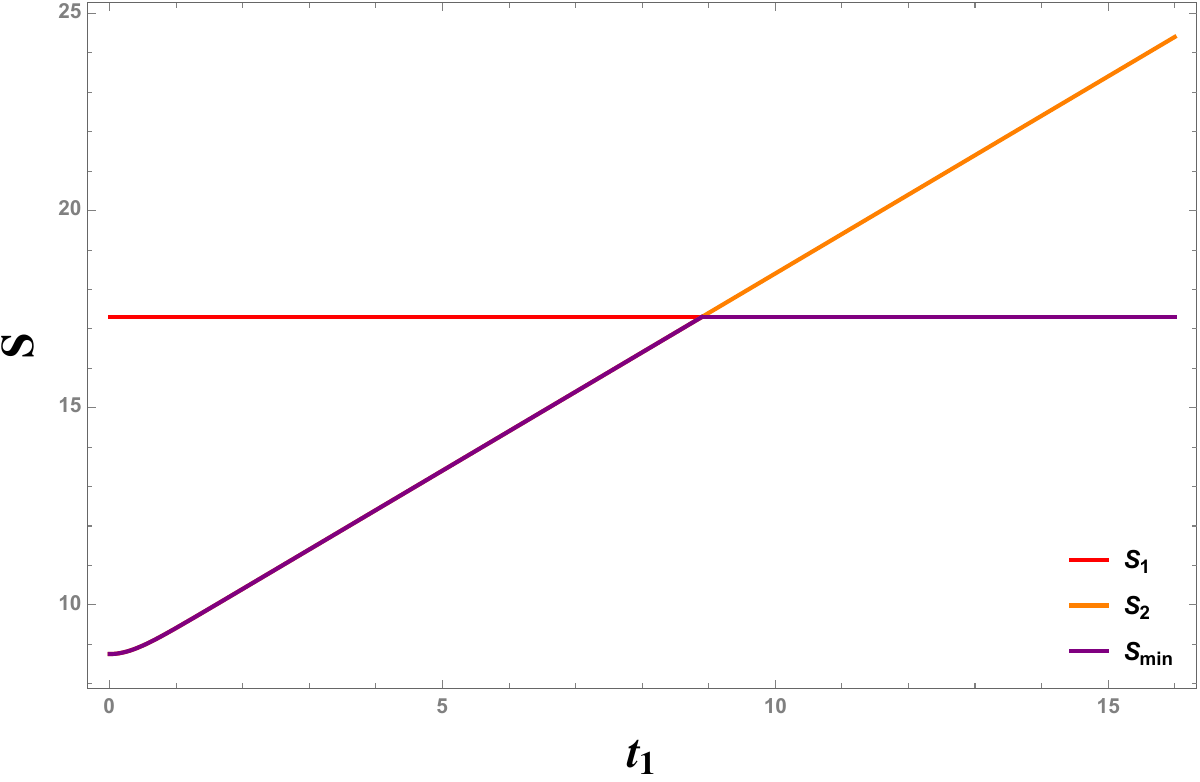}
			\caption{}
			\label{plot-EE-adjacent1}
		\end{subfigure}
		\hfill
		\begin{subfigure}[b]{0.45\linewidth}
			\centering
			\includegraphics[scale=.33]{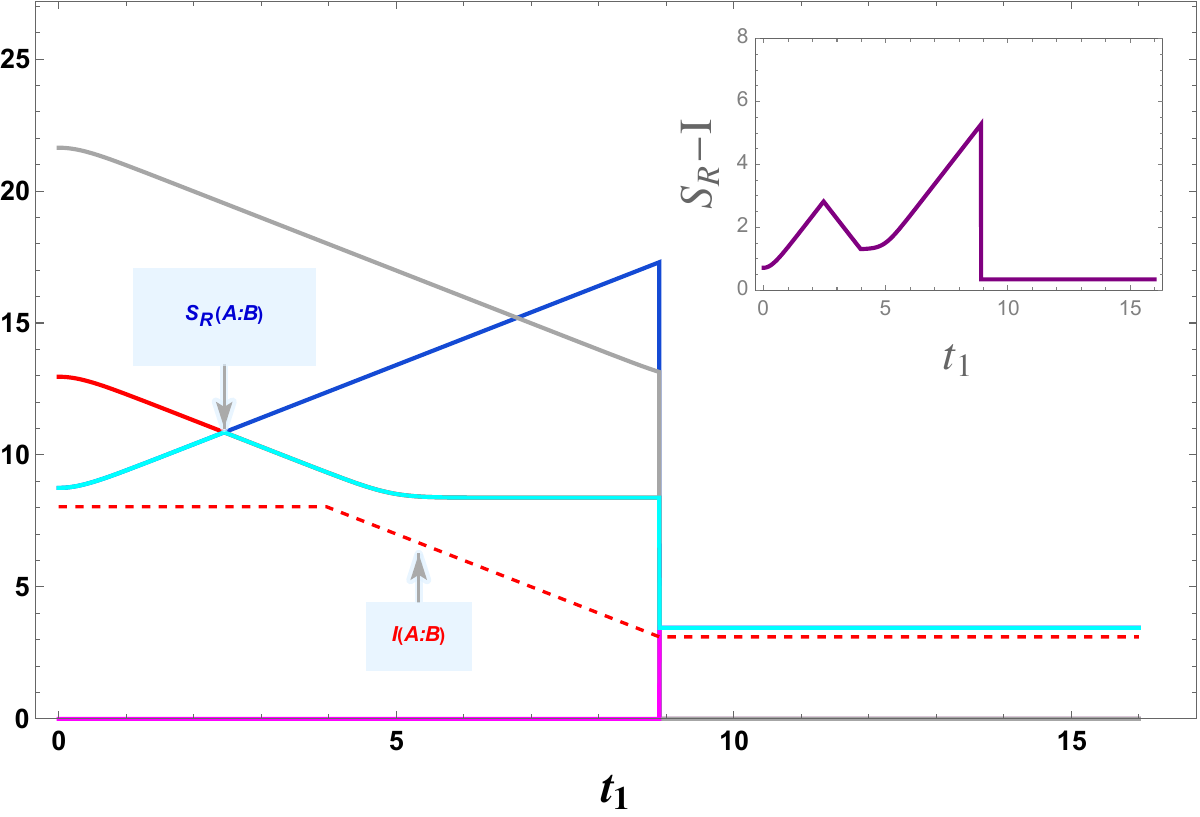}
			\caption{}
			\label{adjacent SR1}
		\end{subfigure}
		\caption{(a) EE for two adjacent subsystems $A \cup B$ vs time graph. Here purple curve indicates the minimum	EE among two phases. (b) Reflected entropy between subsystem $A$ and $B$ as a function of time. Here cyan curve shows minimum $S_R$ and red dashed curve is mutual information. (Both graphs are in units of $c$). The inset graph represents the difference between $S_R$ and mutual information. These plots are obtained with $\ell=1, T=.95, r_H=2 \ell, \epsilon=.001, \epsilon_b=.001, \phi_1=\frac{\pi}{16}, \phi_2=3.2 \pi, \phi_3=9.1 \pi.$ }
	\end{figure}

	For this case also we consider that $w_1$ and $w_3$ are far away from each other, therefore we obtain the EE phase transition between \hyperref[EE-adj2]{phase-II} and \hyperref[EE-adj1]{phase-I}, as shown in \cref{plot-EE-adjacent1}. The transition time for this is given in \cref{adj-EE-trans.-time}.

	We now analyse the time evolution of the reflected entropy across different entanglement entropy (EE) phases, focusing on the scenario where subsystem $A$ is smaller than subsystem $B$.
	At early times, in \hyperref[EE-adj2]{phase-II}, the reflected entropy increases with time as the EWCS is the HM surface and then slowly decreases and and eventually stabilizes till the transition time $T^{\text{adj}}_{E}$ as the EWCS lands on the HM surface corresponding to the points $w_1$ and $w^{\prime}_1$. Finally, in \hyperref[EE-adj1]{phase-I}, it saturates to a constant value. 
	The inset plot indicates that initially the Markov gap is non zero, despite the fact that the bulk EWCS in this phase has no non-trivial boundaries. This once again suggests the need to critically reassess the geometric interpretation of the Markov gap given in \cref{Markov gap}.
	After that this gap increases to a value greater than $\frac{c}{3}\log2$ as the bulk EWCS has one non-trivial boundary in this reflected entropy phase. Finally, in \hyperref[EE-adj1]{phase-I} this gap saturates to the lower bound given in \cref{Markov gap}.

	\subsection{Disjoint subsystems}
	In this subsection, we investigate the time evolution of the reflected entropy between two disjoint subsystems. Here also we observe five different mutual information phases which are given as follows
	\begin{align}
		I(A:B)=\begin{cases}
			\displaystyle&\frac{c}{3}\log\left[\frac{\sinh \left(\frac{r_h \phi_{21}}{2 \ell}\right) \sinh \left(\frac{r_h \phi_{43}}{2 \ell}\right)}{\sinh\left(\frac{r_h \phi_{41}}{2 \ell}\right)\sinh\left(\frac{r_h \phi_{32}}{2 \ell}\right)}\right],\notag\\
			&\frac{c}{3}\log\left[\frac{2 \ell^2 }{r_h\epsilon }\frac{\sinh \left(\frac{r_h \phi_{21}}{2 \ell}\right) \sinh \left(\frac{r_h \phi_{43}}{2\ell}\right)}{\sinh\left(\frac{r_h \phi_{32}}{2 \ell}\right)}\right]-\frac{c}{3}\left(\log \frac{2 \ell^2 \cosh \frac{r_h t_1}{\ell^2}}{r_h \epsilon }+\log \frac{2 r_0 \ell}{r_h \epsilon_b }+\log \sqrt{\frac{1+ T \ell }{1-T \ell }}\right),\notag\\
			&\frac{c}{3}\log\left[\frac{\sinh \left(\frac{r_h \phi_{21}}{2 \ell}\right)}{\sinh\left(\frac{r_h \phi_{32}}{2 \ell}\right)}\right],\\
			&\frac{c}{3}\log\left[\frac{\sinh \left(\frac{r_h \phi_{43}}{2 \ell}\right)}{\sinh\left(\frac{r_h \phi_{32}}{2 \ell}\right)}\right],\notag\\
			&\frac{c}{3}\left(\log \frac{2 \ell^2 \cosh \frac{r_h t_1}{\ell^2}}{r_h \epsilon }+\log \frac{2 r_0 \ell}{r_h \epsilon_b }+\log \sqrt{\frac{1+ T \ell }{1-T \ell }}\right)-\frac{c}{3}\log \left[\frac{2 \ell^2}{r_h \epsilon}\sinh\left(\frac{r_h \phi_{32}}{2 \ell}\right)\right].
		\end{cases}
	\end{align}

	\subsection*{Case-I}
	
	\begin{figure}[ht]
		\centering
		\begin{subfigure}[b]{0.45\linewidth}
			\centering
			\includegraphics[scale=.35]{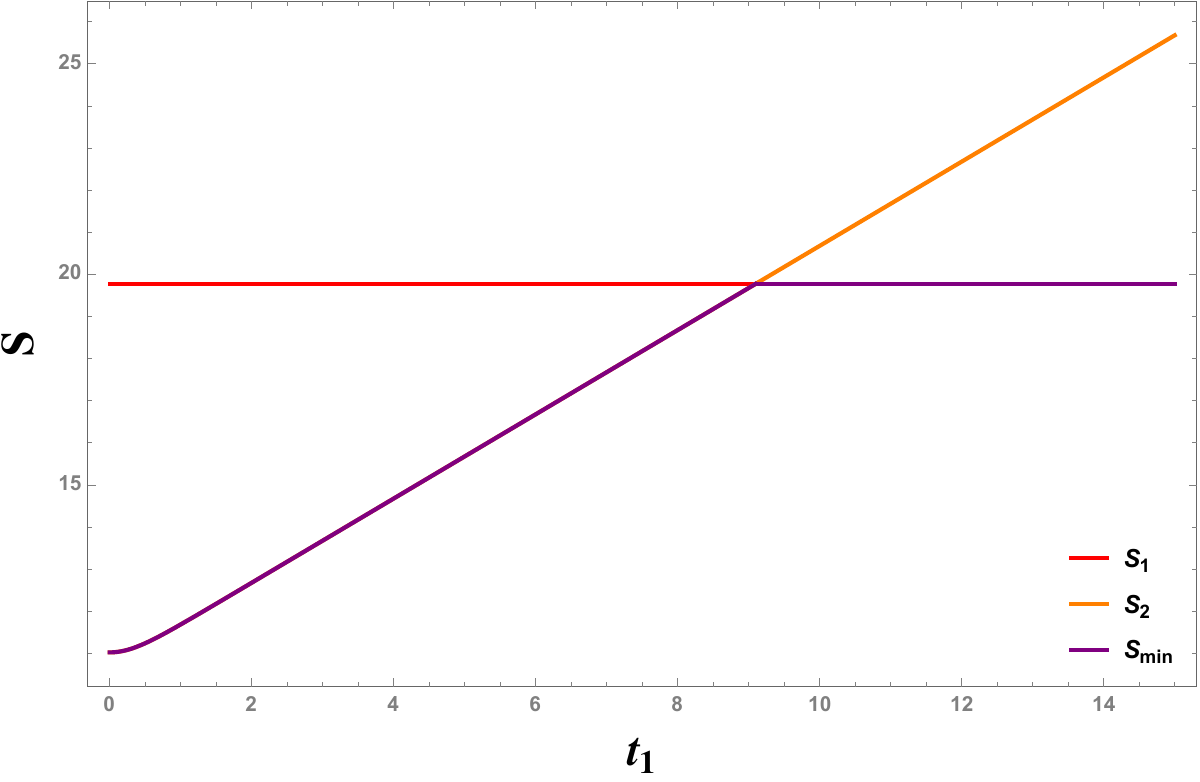}
			\caption{}
			\label{dis plot(EE-disjoint1)}
		\end{subfigure}
		\hfill
		\begin{subfigure}[b]{0.45\linewidth}
			\centering
			\includegraphics[scale=.24]{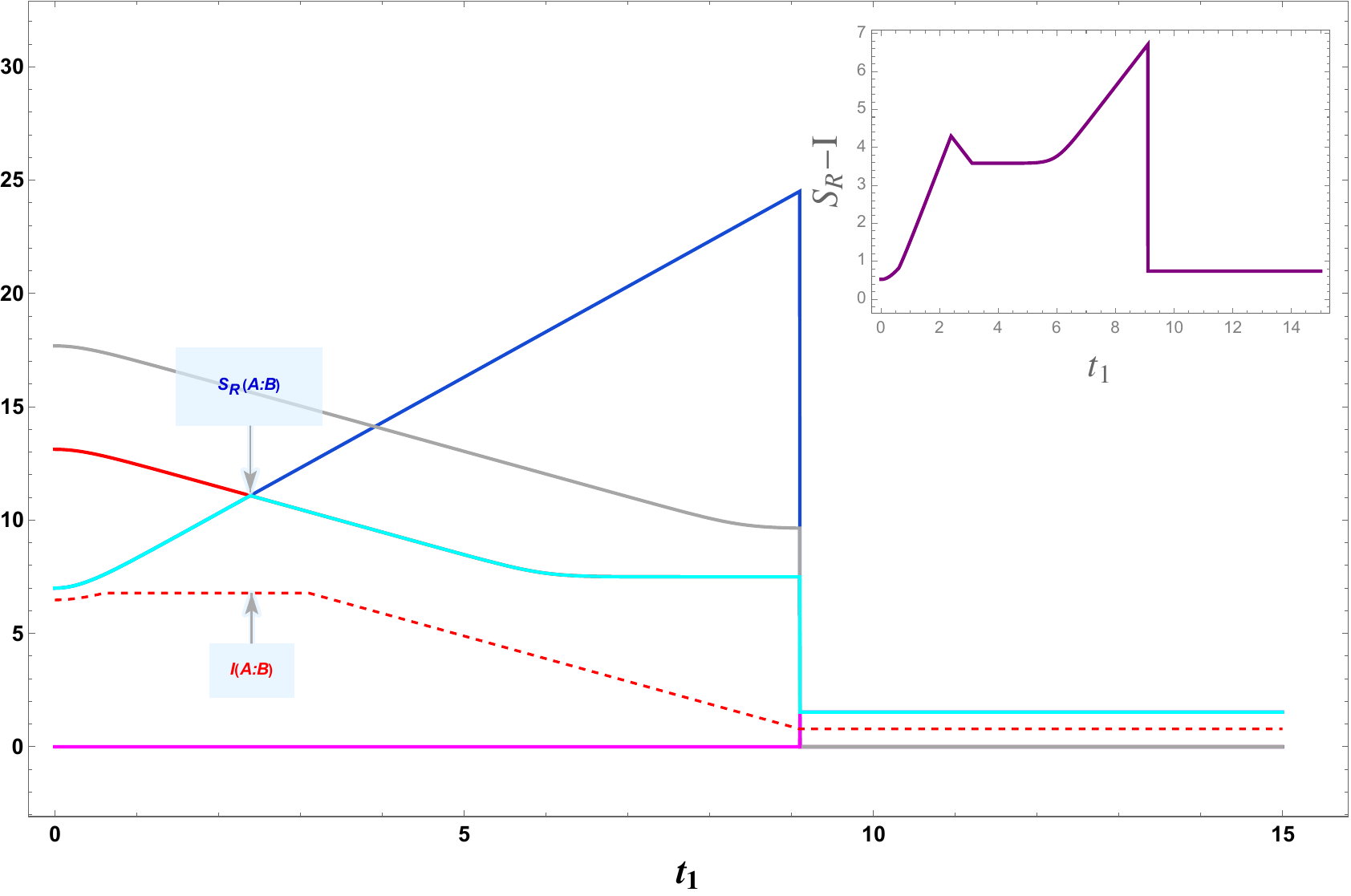}
			\caption{}
			\label{fig:disjoint SR 1}
		\end{subfigure}
		\caption{(a) EE for two disjoint subsystems $A \cup B$ vs time graph. Here purple curve indicates the minimum	EE among two phases. (b) Reflected entropy between subsystem $A$ and $B$ as a function of time. Here cyan curve shows minimum $S_R$ and red dashed curve is mutual information. (Both graphs are in units of $c$). The inset graph represents the difference between $S_R$ and mutual information. These plots are obtained with $\ell=1, T=.95, r_H=2 \ell, \epsilon=.001, \epsilon_b=.001, \phi_1=\frac{\pi}{3}, \phi_2=4.12 \pi, \phi_3=4.15 \pi, \phi_4=9.5 \pi.$ }
	\end{figure}
	The entanglement entropy phase transition between \hyperref[EE-dis-phase-2]{phase-II} and \hyperref[EE-dis-phase-1]{phase-I} may be obtained by considering that subsystems $A$ and $B$ are far away from each other. This EE phase transition is shown in \cref{dis plot(EE-disjoint1)} and the transition time is given as
	\begin{align}\label{dis-EE-trans.-time}
		T^{\text{disj}}_{E}=\frac{\ell^2}{r_h}	\cosh ^{-1}\left(\frac{\epsilon_b(1- T \ell ) \sinh \frac{r_h \phi_{41}}{2 \ell}}{2 \ell^2}\right).
	\end{align}
	We now investigate the time evolution of the reflected entropy in these EE phases, considering that the subsystem $A$ is smaller than the subsystem $B$. The reflected entropy phase transition is depicted in \cref{fig:disjoint SR 1}. Initially in \hyperref[EE-dis-phase-2]{phase-II} , the reflected entropy increases with time as the bulk EWCS is given by the HM surface, then slowly starts decreasing and remain constant until the transition time $T^{\text{disj}}_{E}$ as the bulk EWCS lands on the HM surface corresponding to the points $w_1$ and $w^{\prime}_1$. Finally, in \hyperref[EE-dis-phase-1]{phase-I} it saturates to a constant value. 
	From the inset plot, we observe that initially the Markov gap is greater than $\frac{c} {3}\log 2$ as the bulk EWCS has one non-trivial boundary and after that this gap increases to $ \frac{2c} {3}\log 2$ due to two non-trivial boundaries of the bulk EWCS. Finally, after the transition time $T^{\text{disj}}_{E}$, the Markov gap saturates to the lower bound given in \cref{Markov gap}.

	\subsection*{Case-II}
	\begin{figure}[ht]
		\centering
		\begin{subfigure}[b]{0.45\linewidth}
			\centering
			\includegraphics[scale=.35]{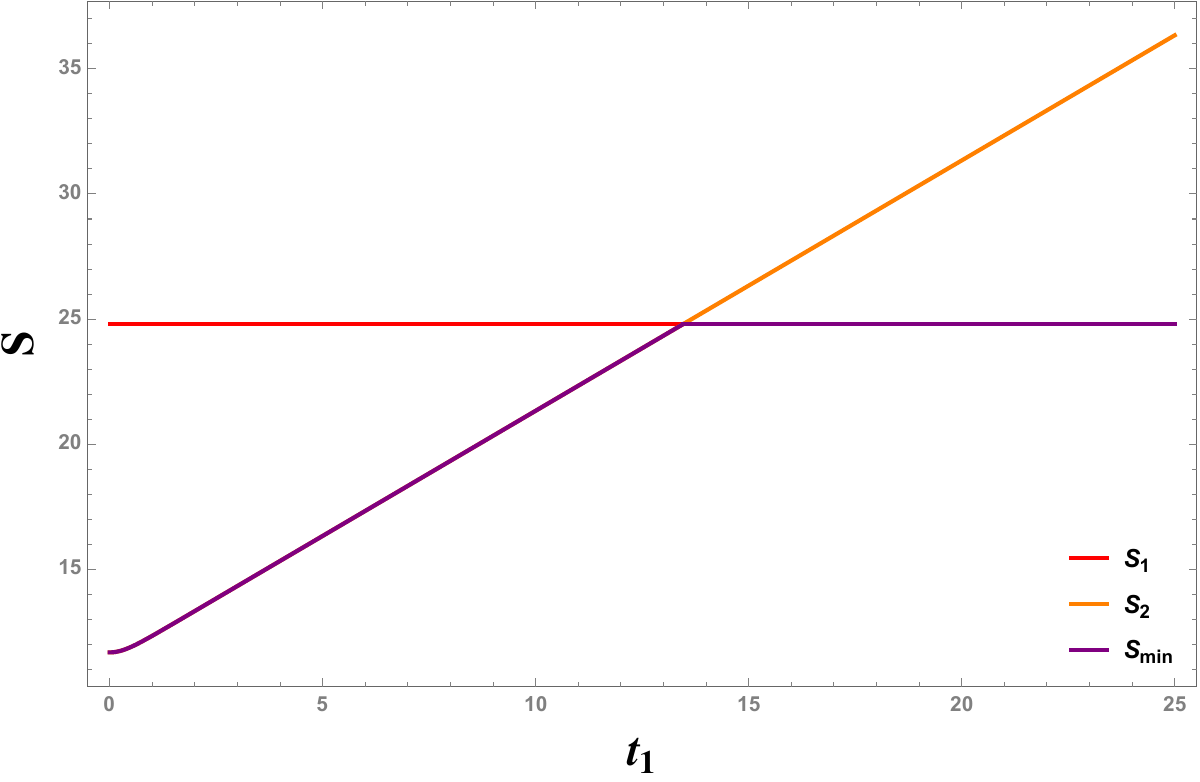}
			\caption{}
			\label{dis plot(EE-disjoint2)}
		\end{subfigure}
		\hfill
		\begin{subfigure}[b]{0.45\linewidth}
			\centering
			\includegraphics[scale=.33]{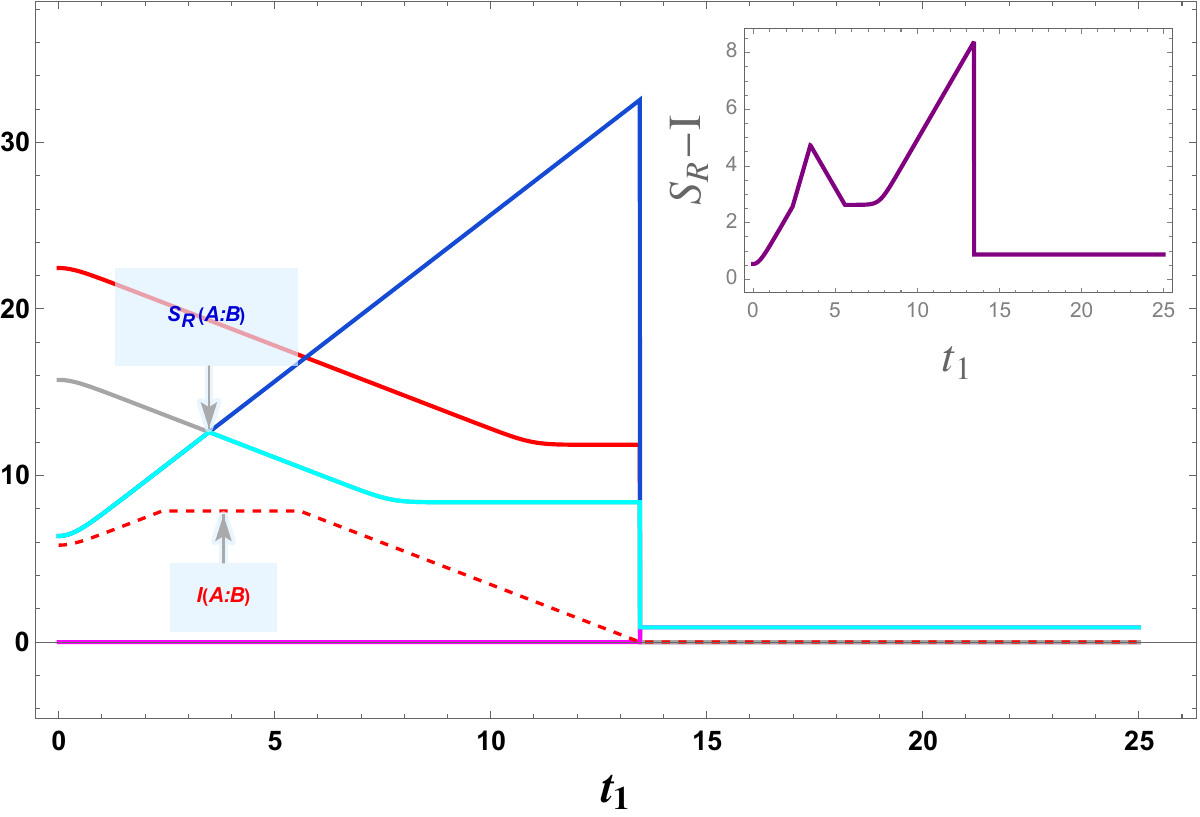}
			\caption{}
			\label{disjoint SR2 }
		\end{subfigure}
		\caption{(a) EE for two disjoint subsystems $A \cup B$ vs time graph. Here purple curve indicates the minimum	EE among two phases. (b) Reflected entropy between subsystem $A$ and $B$ as a function of time. Here cyan curve shows minimum $S_R$ and red dashed curve is mutual information. (Both graphs are in units of $c$). The inset graph corresponds to the geometric Markov gap. These plots are obtained with $\ell=1, T=.95, r_H=2 \ell, \epsilon=.001, \epsilon_b=.001, \phi_1=\frac{\pi}{16}, \phi_2=6.99 \pi, \phi_3=7.1 \pi, \phi_4=12 \pi.$}
	\end{figure}
	
	For this case also we consider that both the subsystems are far away from each other, therefore we obtain the EE phase transition between \hyperref[EE-dis-phase-2]{phase-II} and \hyperref[EE-dis-phase-1]{phase-I}, as shown in \cref{plot-EE-adjacent1}. The transition time for this is given in \cref{dis-EE-trans.-time}. 
	
	Now we illustrate the variation of the reflected entropy over time, by considering that the subsystem $B$ is smaller than the subsystem $A$. Initially, the reflected entropy increases with time as the bulk EWCS is the HM surface, 
	However, after a certain time, it gradually decreases and and remain constant until the transition time $T^{\text{disj}}_{E}$ as the bulk EWCS lands on the HM surface associated with the points $w_3$ and $w^{\prime}_3$. Finally, the reflected entropy saturates to a constant value in \hyperref[EE-dis-phase-1]{phase-I}. 
	The inset plot shows that initially the Markov gap is always greater than $\frac{c}{3}\log 2$ since there is one non-trivial boundary for the bulk EWCS phase and after that it increases to $\frac{2c}{3}\log 2$ due to the two non-trivial boundaries of the bulk EWCS. Finally in \hyperref[EE-dis-phase-1]{phase-I}, this gap saturates to the lower bound mentioned in \cref{Markov gap}.

	\section{Summary and conclusions}\label{sec:conc}
	In this work, we have presented a detailed investigation into the structure of reflected entropy and its associated phases in the context of a braneworld cosmology, which is described by an eternal BTZ black hole truncated by an end-of-the-world (EOW) brane. The setup offers a lower-dimensional effective description in terms of a braneworld cosmology coupled to a BCFT$_2$ \cite{Wang:2021xih}. Our primary focus has been the computation of reflected entropy for two adjacent and disjoint subsystems. We employed two distinct prescriptions -- the island prescription and the defect extremal surface (DES) prescription. The agreement between these two approaches in the large central charge limit is remarkable, confirming that both prescriptions provide consistent results for computing reflected entropy in the braneworld cosmology.
	
	A significant part of our study involves analyzing the time evolution of reflected entropy. We observed that reflected entropy evolves in a non-trivial way, reflecting a rich phase structure depending on the size and configuration of the subsystems. The behavior of reflected entropy varies across different phases, and we classified these into distinct regimes. For adjacent subsystems, we found two distinct entanglement entropy phases, with the reflected entropy showing different dependencies based on the location and relative sizes of the subsystems under consideration. In the case of disjoint subsystems, the phases were also characterized by the separation of the subsystems, with different reflected entropy phases emerging depending on the relative size of the subsystems and their separation in the boundary.
	
	In addition to the reflected entropy, we have also examined the holographic mutual information. Furthermore, we found that the Markov gap, an indicator of tripartite entanglement, persists even in cases where the EWCS boundaries are trivial. This is an important observation, as the non-zero Markov gap indicates that there are subtle correlations in the mixed state structure, even in situations where the entanglement wedges do not exhibit nontrivial boundaries.
	
	One of the novel aspects of our analysis is the identification of extremal surfaces that do not cross the horizon, yet still probe regions of the black hole interior. This offers a new perspective on how quantum extremal surfaces can provide indirect probes of the black hole's interior without requiring direct access to the black hole or cosmological horizon. The existence of such surfaces suggests that quantum extremal surfaces, in particular the extremal EWCS, may serve as effective tools for probing the quantum structure of the black hole interior. These surfaces provide a way to explore the quantum features behind the horizon, offering an alternative approach to understanding black hole information paradoxes.
	
	Additionally, our work raises interesting possibilities for extending the study of mixed state entanglement to more general cosmological settings, including higher-dimensional spacetimes and different configurations of EOW branes. While this study focuses on a lower-dimensional braneworld model, the principles and methods employed could be applied to more complex systems with higher-dimensional black holes and cosmological setups. This could lead to new insights into the role of entanglement and quantum correlations in cosmological spacetimes, and further exploration could shed light on the potential interplay between cosmological horizons, quantum extremal surfaces, and holography.
	
	In summary, this work provides new insights into the nature of mixed state entanglement in braneworld cosmologies, with a particular focus on reflected entropy and holographic mutual information. We have explored different entanglement entropy phases and shown how the defect extremal surface and island prescriptions can be applied to compute reflected entropy in such cosmologies. Our findings suggest that quantum extremal surfaces may play a key role in understanding the quantum structure of the black hole interior and cosmological spacetimes more broadly.

	\appendix
	\section{BTZ in Kruskal coordinates}\label{AppA}
\begin{figure}[ht]
		\centering
		\includegraphics[width=0.55\textwidth]{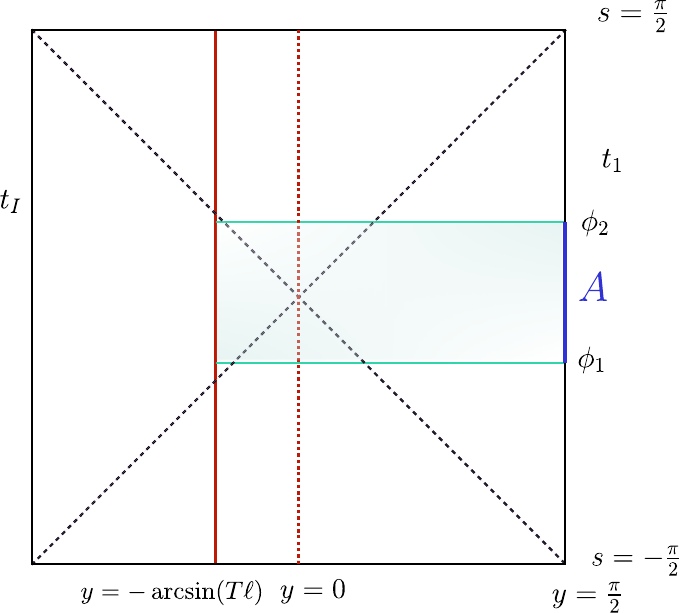}
		\caption{Penrose diagram of the BTZ black hole in Kruskal coordinates $(s,y)$. Here green lines are the RT surface for subsystem $A$. The solid red line denotes the EOW brane with tension $T$ while the dashed red line represents the brane with zero tension. }
		\label{fig:BTZ_Penrose}
	\end{figure}

	In this appendix, we provide some details on the metric \eqref{BTZ-Kruskal} following \cite{Cooper:2018cmb,Wang:2021xih}. We first go to the Kruskal coordinates by defining \cite{Cooper:2018cmb}
	\begin{align}
		r=r_H\frac{1-u v}{1+u v}~~,~~t=\frac{\ell^2}{2r_H}\log\left(-\frac{u}{v}\right)\label{Kruskal-map-1}
	\end{align}
	In these coordinates, the BTZ metric \eqref{BTZ-metric} takes the form
	\begin{align}
		\d s^2=-4 \ell^2 \frac{\d u \d v}{(1+u v)^2}+r_H^2 \frac{(1-u v)^2}{(1+u v)^2}\d\phi ^2 
	\end{align}
	These coordinates cover the full maximally extended BTZ black hole spacetime. In the second asymptotic region, the Schwarzschild coordinates are given by \eqref{Kruskal-map-1} with $u$ and $v$ interchanged. In particular, the EOW brane trajectory in the Lorentzian signature is given as
	\begin{align}
		\frac{v-u}{\sqrt{(1+u^2)(1+v^2)}}=T\ell\,.
	\end{align}
	To obtain the metric \eqref{BTZ-Kruskal}, one further defines
	\begin{align}
		u=\tan\alpha~~,~~v=\tan\beta\,,
	\end{align}
	along with $s=\alpha+\beta$ and $y=\alpha-\beta$. The final form of the coordinate transformation from $(r,t)$ to $(s,y)$ coordinates is then given as \cite{Wang:2021xih}
	\begin{align}
		r&=r_H\frac{1-\tan \left(\frac{s+y}{2}\right) \tan \left(\frac{s-y}{2}\right)}{1+\tan \left(\frac{s+y}{2}\right) \tan \left(\frac{s-y}{2}\right)}=r_H\frac{\cos s}{\cos y}\,,\notag\\
		t&=\frac{\ell^2}{2r_H}\log\left[\frac{\tan \left(\frac{s+y}{2}\right)}{\tan \left(\frac{s-y}{2}\right)}\right]=\frac{\ell^2}{2r_H}\log\left(\frac{\sin y+\sin s}{\sin y-\sin s}\right)\,.
	\end{align}
	The range of these coordinates are given by $-\frac{\pi}{2}\leq s,y\leq\frac{\pi}{2}$. The horizons are given by the asymptotics $y=\pm s$, while the future and past singularities are given by $s=\pm\frac{\pi}{2}$. The two asymptotic boundaries are reached at $y=\pm\frac{\pi}{2}$. However, due to the presence of the EOW brane, we only have access to the right boundary $y=\frac{\pi}{2}$. The Lorentzian trajectory of the EOW brane takes the particularly simple form
	\begin{align}
		y=-\arcsin(T\ell)
	\end{align}
	The Penrose diagram of the maximally extended BTZ black hole is depicted in \cref{fig:BTZ_Penrose}. 
	\subsection{Length of extremal surfaces}
	Generic spacelike geodesics in this spacetime are given by \cite{Cooper:2018cmb}
	\begin{align}
		\sin\left(s_B-s_0\right)\sin y=\sin\left(s-s_0\right)
	\end{align}
	where the geodesic ends on the asymptotic boundary at $s=s_B$ and passes through $s=s_0$ at $y=0$. We first consider the extremal surface which crosses the horizon and ends on the EOW brane. As discussed in \cite{Rozali:2019day}, in the Kruskal geometry the extremal condition reduces to the geometric constraint that the geodesics are normal to the brane. This leads to the relatively simple class of geodesics 
	\begin{align}
		s=s_0\,.
	\end{align}
	These geodesics are depicted by the green dashed lines in \cref{fig:BTZ_Penrose}.
	
	To compute the length of such extremal surfaces, we utilize the inverse transformations
	\begin{align}
		s=\arctan\left[\frac{\sqrt{r^2-r_H^2}}{r}\sinh\left(\frac{r_H t}{\ell^2}\right)\right]~~,~~y=\arctan\left[\frac{\sqrt{r^2-r_H^2}}{r_H}\cosh\left(\frac{r_H t}{\ell^2}\right)\right]\,.\label{inverse-transformations}
	\end{align}
	For a geodesic endpoint $(\phi_1,\frac{t_1}{\ell})$ on the boundary, we may use a regulator surface at
	\begin{align}
		y_\textrm{max}=\arctan\left[\frac{\ell^2}{\epsilon r_H}\cosh\left(\frac{r_H t_1}{\ell^2}\right)\right]
	\end{align}
	corresponding to $r_\textrm{max}=\frac{\ell^2}{\epsilon}$. Therefore, we may find the length of the extremal surface extending to the EOW brane as follows
	\begin{align}
		\mathcal{L}_\textrm{HM}&=\ell\int_{-\arcsin(T\ell)}^{y_\textrm{max}}\frac{\d y}{\cos y}\notag\\&=\ell\,\log\left[\frac{2\ell^2}{\epsilon r_H}\cosh\left(\frac{r_H t_1}{\ell^2}\right)\right]+\ell\,\textrm{arctanh}(T\ell)\notag\\&=\ell\,\log\left[\frac{\beta}{\pi\epsilon}\cosh\left(\frac{2\pi t_1}{\beta}\right)\right]+\ell\,\log\sqrt{\frac{1+T\ell}{1-T\ell}}
	\end{align}
	
	Next, we consider the dome-type extremal surface anchored on two boundary points $(\phi_1,\frac{t_1}{\ell})$ and $(\phi_2,\frac{t_1}{\ell})$. To compute the length of this geodesic, we utilize the embedding coordinate formalism. For the Kruskal metric \eqref{BTZ-Kruskal}, the embedding coordinates are easily found to be \cite{Shenker:2013pqa,Doi:2023zaf}
	\begin{align}
		X^0&=\ell\frac{u+v}{1+u v}=\ell\,\sec y\sin s\,,\notag\\
		X^1&=\ell\frac{1-u v}{1+u v}\cosh\left(\frac{r_H t}{\ell^2}\right)=\ell\,\sec y\cos s\cosh\left(\frac{r_H t}{\ell^2}\right)\,,\notag\\
		X^2&=\ell\frac{1-u v}{1+u v}\sinh\left(\frac{r_H t}{\ell^2}\right)=\ell\,\sec y\cos s\sinh\left(\frac{r_H t}{\ell^2}\right)\,,\notag\\
		X^3&=\ell\frac{v-u}{1+u v}=-\ell\,\tan y\,.
	\end{align}
	The length of a geodesic connecting two bulk points $(s_1,y_1,\phi_1)$ and $(s_2,y_2,\phi_2)$ may be comprehensively obtained utilizing the formula
	\begin{align}
		\mathcal{L}_{12}&=\ell\,\textrm{arccosh}\left(-\frac{1}{\ell^2}X[s_1,y_1,\phi_1]\cdot X[s_2,y_2,\phi_2]\right)\notag\\
		&=\ell\,\textrm{arccosh}\left[\sec y_1\sec y_2\left(\cos s_1\cos s_2\cosh\left(\frac{r_H(\phi_1-\phi_2)}{\ell}\right)+\sin s_1\sin s_2\right)-\tan y_1\tan y_2\right]\label{s-y-geodesic-length}
	\end{align}
	From \cref{inverse-transformations}, it is easy to verify that for the boundary points $(\phi_1,\frac{t_1}{\ell})$ and $(\phi_2,\frac{t_1}{\ell})$, 
	\begin{align}\label{s-y-r-t}
		s_1=s_2=\arctan\left[\sinh\left(\frac{r_H t_1}{\ell^2}\right)\right]~~,~~y_{\text{max}}=y_1=y_2=\arctan\left[\frac{\ell^2}{\epsilon r_H}\sinh\left(\frac{r_H t_1}{\ell^2}\right)\right]
	\end{align}
	Hence the length of the minimal surface joining these two points may be readily obtained from \cref{s-y-geodesic-length} as
	\begin{align}
		\mathcal{L}_\textrm{RT}=\ell\,\log\left[\frac{2\ell^2}{\epsilon r_H}\sinh\frac{r_H(\phi_2-\phi_1)}{2\ell}\right]=\ell\,\log\left[\frac{\beta}{\pi\epsilon}\sinh\left(\frac{\pi(\phi_2-\phi_1)}{\beta}\right)\right]
	\end{align}
	
	\subsection{EWCS in Kruskal coordinates}
	We now illustrate the computation of the bulk EWCS in the Kruskal coordinates through two examples.
	\subsubsection{Adjacent subsystems: EWCS lands on the HM surface}\label{geodesic-length-s-y-adj2(i)}
	The geodesic length of the curve $\Sigma_{AB}$, shown as green colour in \cref{fig:adj2(i)}, may also obtain by using the Kruskal-like coordinates $(s,y)$. 
	Utilizing the end points $(s_2,y_{\text{max}},\phi_2)$ and $(\tilde{s},\tilde{y},\tilde{\phi})$ of the curve $\Sigma_{AB}$ in \cref{s-y-geodesic-length}, the geodesic length of may be written as
	\begin{align}\label{length-adj2(i)(s-y)}
		\mathcal{L}({\Sigma_{AB}})
		=\ell\,\textrm{arccosh}\Bigg[\sec y_{\text{max}}\sec \tilde{y} \left(\cos s_2\cos \tilde{s} \cosh\left(\frac{r_H(\phi_2-\tilde{\phi})}{\ell}\right)+\sin s_2\sin \tilde{s}\right)&\notag\\-\tan y_{\text{max}}\tan \tilde{y}\Bigg]&,
	\end{align}
	where from the geometry of the geodesic in Kruskal coordinate we may find that $ \tilde{s}=s_1$ and $\tilde{\phi}=\phi_{1}$ and $\tilde{y}$ is an arbitrary point on the HM surface corresponding to the point $w_1$ and $w^{\prime}_1$. Using these arguments in the above equation and then extremizing the resulting expression over $\tilde{y}$, we get the extremum value of $\tilde{y}$ as
	\begin{align}
		\tilde{y}=\csc^{-1}\left[\csc y_{\text{max}} \left(\cos s_2\cos s_1 \cosh\left(\frac{r_H(\phi_2-\phi_1)}{\ell}\right)+\sin s_2\sin s_1\right)\right]
	\end{align}
	Substituting the extremized value of $\tilde{y}$ in \cref{length-adj2(i)(s-y)} and then transforming $s_1$, $s_2$ and $y_{\text{max}}$ back to the original coordinates using \cref{s-y-r-t}, the geodesic length of the curve $\Sigma_{AB}$ may be obtained and is exactly equal to \cref{Geodesic-length-SR-adj2(i)}.
	
	\subsubsection{ Disjoint subsystems: EWCS lands on the EOW brane}\label{geodesic-length-s-y-dis2(i)}
	The geodesic length of the curve $\Sigma_{AB}$, depicted as green colour in \cref{fig:dis-2(ii)} may also be computed by utilizing the Kruskal-like coordinates $(s,y)$ as 
	\begin{align}\label{SR-dis-2(ii)-s-y}
		\mathcal{L}({\Sigma_{AB}})&=\ell\int_{-\arcsin(T\ell)}^{y_{d}}\frac{\d y}{\cos y},
	\end{align}
	where $y_{d}$ is a point on the dome-type RT surface.

	The geodesic equation of dome-type RT surface anchored on two boundary points $(\phi_1,\frac{t_1}{\ell})$ and $(\phi_2,\frac{t_1}{\ell})$ is given as
	\begin{align}
		r(\phi)= r_h \frac{\cosh\frac{r_h(\phi_3-\phi_2)}{2\ell}}{\sqrt{\sinh\frac{r_h(\phi_3-\phi)}{\ell}\sinh\frac{r_h(\phi-\phi_2)}{\ell}}},
	\end{align}
	where $(r,\phi)$ corresponds to a generic point of the RT surface. 
	To find the minimal length of the curve $\Sigma_{AB}$, it is necessary that the curve must starts from the tip of the dome-type RT surface which may be determined by extremizing the above expression over $\phi$. By extremizing we get $\phi=\frac{\phi_2+\phi_3}{2}$ and substituting this we may obtain the coordinate of the highest point on the dome-type RT surface as
	\begin{align}
		r_d=  r_h \coth\frac{r_h(\phi_3-\phi_2)}{2 \ell}.
	\end{align}
	Putting this value in \cref{inverse-transformations}, we get
	\begin{align}
		y_d=\textrm{arctan}\left[\frac{\cosh\frac{r_h t_1}{\ell^2}}{\sinh\frac{r_h(\phi_3-\phi_2)}{2 \ell}}\right].
	\end{align}
	Now substituting $y_d$ in \cref{SR-dis-2(ii)-s-y} and performing the integration, we can obtain the length of the curve $\Sigma_{AB}$ which is identically equal to \cref{area_dis2(ii)}.
	
	\section{Minimal length between two extremal curves}\label{appB}
	In this appendix, we collect some important results from \cite{Kusuki:2019evw} for the minimal length between two extremal curves in asymptotically AdS$_3$ spacetimes, which will be relevant to our discussion. Consider two disjoint subsystems, $A=[X_1,X_2]$ and $B=[X_3,X_4]$ on the boundary of an asymptotically AdS$_3$ spacetime, written in the embedding coordinates \eqref{embedding-coordinates}. For a connected entanglement wedge, the extremal surfaces computing the entanglement entropy of $A\cup B$ are given as \cite{Kusuki:2019evw}
	\begin{align}
		X_{14}^A(\lambda)=\frac{X_1^A e^{-\lambda}+X_4^A e^\lambda}{\sqrt{2\zeta_{14}}}~~,~~X_{23}^A(\bar\lambda)=\frac{X_2^A e^{-\bar\lambda}+X_3^A e^{\bar\lambda}}{\sqrt{2\zeta_{23}}}\,,
	\end{align}
	where $\zeta_{ij}=-X_i\cdot X_j$ and $(\lambda,\bar\lambda)$ are real affine parameters on the extremal surfaces. The EWCS corresponds to a geodesic curve of minimal length joining these two curves. As described in \cite{Kusuki:2019evw}, one may reformulate the problem of finding the EWCS as an optimization problem of the length of this curve over the affine parameters $(\lambda,\bar\lambda)$ as follows
	\begin{align}
		\mathcal{L}(\lambda,\bar\lambda)&\equiv \mathcal{L}\left(X_{14}(\lambda)\cdot X_{23}(\bar\lambda)\right)\notag\\
		&=\cosh^{-1}\left[\frac{\zeta_{12}e^{-\lambda-\bar\lambda}+\zeta_{13}e^{-\lambda+\bar\lambda}+\zeta_{24}e^{\lambda-\bar\lambda}+\zeta_{34}e^{\lambda+\bar\lambda}}{2\sqrt{\zeta_{14}\zeta_{23}}}\right]
	\end{align}
	The optimized values are given by
	\begin{align}
		\lambda_\star=\frac{1}{4}\log\left(\frac{\zeta_{12}\zeta_{14}}{\zeta_{24}\zeta_{34}}\right)~~,~~\bar\lambda_\star=\frac{1}{4}\log\left(\frac{\zeta_{12}\zeta_{34}}{\zeta_{14}\zeta_{34}}\right)\,,
	\end{align}
	and the EWCS is obtained from \eqref{EWCS-Dis}.
	
	As an illustration, utilizing the embedding coordinates \eqref{embedding-r} for the BTZ black hole, we may obtain the location of the extremal point on the HM surface as follows
	\begin{align}
		\ell^2\frac{r^2}{r_h^2}&=\left(X^1_{14}\left(\lambda_\star\right)\right)^2-\left(X^3_{14}\left(\lambda_\star\right)\right)^2\\
		\tanh\left(\frac{r_h\phi}{\ell}\right)&=\frac{X_{14}^3\left(\lambda_\star\right)}{X_{14}^1\left(\lambda_\star\right)}=\frac{X_1^3+X_4^3e^{2\lambda^{\star}}}{X_1^1+X_4^1e^{2\lambda^{\star}}}=\tanh\left(\frac{r_h\phi_1}{\ell}\right)\\
		\tanh\left(\frac{r_h t}{\ell^2}\right)&=\frac{X_{14}^0\left(\lambda_\star\right)}{X_{14}^2\left(\lambda_\star\right)}=\frac{X_1^0+X_4^0e^{2\lambda^{\star}}}{X_1^2+X_4^2e^{2\lambda^{\star}}}=\tanh\lambda^{\star}\tanh\left(\frac{r_h t_1}{\ell^2}\right)\,,
	\end{align}
	where the optimal value of the affine parameter is given as
	\begin{align}
		e^{2\lambda_\star}=\frac{\cosh\left(\frac{r_h t_1}{\ell^2}\right)\sinh\left(\frac{r_h\phi_{21}}{\ell}\right)}{\sqrt{\left[\cosh^2\left(\frac{r_h t_1}{\ell^2}\right)+\sinh^2\left(\frac{r_h\phi_{21}}{\ell}\right)\right]\left[\cosh^2\left(\frac{r_h t_1}{\ell^2}\right)+\sinh^2\left(\frac{r_h\phi_{31}}{\ell}\right)\right]}}
	\end{align}
	In particular, the radial location of the extremal point on the HM surface is given by
	\begin{align}
		r=r_h\sqrt{1+\sinh^2\lambda_\star\sech^2\left(\frac{r_h t_1}{\ell^2}\right)}\,.
	\end{align}
	For $t_1=0$, we have 
	\begin{align}
		r^2(t_1=0)=\frac{r_h^2}{8}\frac{\left[\cosh \left(\frac{r_h(\phi_{21}+\phi_{31})}{2\ell}\right)+2 \sinh \left(\frac{r_h\phi_{21}}{2\ell}\right)+\cosh \left(\frac{r_h\phi_{32}}{2\ell}\right)\right]^2}{\sinh\left(\frac{r_h\phi_{21}}{\ell}\right) \cosh\left(\frac{r_h\phi_{31}}{2\ell}\right)}>r_h^2
	\end{align}
	Hence, the EWCS landing on the HM surface does not always probe behind the horizon, similar to the case with adjacent subsystems (cf. \cref{EW-on-HM-r-adj}). As earlier, it may be shown that this surface never crosses the horizon and ends on the asymptotic boundary.

	\bibliographystyle{utphys}
	
	\bibliography{ref}
	
\end{document}